\documentclass[%
 reprint,
nofootinbib,
 amsmath,amssymb,
 aps,
]{revtex4-1}

\usepackage{graphicx}
\usepackage{dcolumn}
\usepackage{bm}
\usepackage{hyperref}
\usepackage[mathlines]{lineno}

\usepackage[T1]{fontenc} 

\usepackage{extarrows} 
\usepackage{amsmath}
\usepackage{graphicx}
\usepackage{dcolumn}
\usepackage{bm}
\usepackage{wasysym}
\usepackage{verbatim}
\usepackage{color}
\usepackage{mathtools,slashed}
\usepackage{cleveref}
\usepackage{subcaption}

\newcommand{\Beq}{\begin{equation}\begin{aligned}}
\newcommand{\Eeq}{\end{aligned}\end{equation}}

\newcommand{\p}{\partial}

\newcommand{\mH}{\mathcal{H}}
\newcommand{\xa}{\xi_{A}}
\newcommand{\xpi}{\xi_{\varphi}}
\newcommand{\an}{\quad \textmd{and} \quad }

\newcommand{\I}{{\bf{I}}_2}
\newcommand{\II}{{\rm{I}_4}}
\newcommand{\III}{{\rm{I}_8}}
\newcommand{\bea}{\begin{eqnarray}}
\newcommand{\eea}{\end{eqnarray}}

\newcommand{\ga}{g_A}

\newcommand{\xp}{\xi_{A}}
\newcommand{\sD}{\slashed{D}}

\newcommand{\x}{\tilde{\tau}}

\newcommand{\bk}{\boldsymbol{k}}
\newcommand{\bx}{\boldsymbol{x}}

\usepackage{bm,upgreek}

\begin{document}


\title{Production and Backreaction of Fermions from Axion-$SU(2)$ Gauge Fields during Inflation}

\author{Leila Mirzagholi}
 \altaffiliation[lmirz@MPA-Garching.MPG.DE]{}
\author{Azadeh Maleknejad}%
 \email{amalek@MPA-Garching.MPG.DE}
 \author{Kaloian D. Lozanov}
 \email{klozanov@MPA-Garching.MPG.DE}
\affiliation{Max-Planck-Institute for Astrophysics, Karl-Schwarzschild-Str.1, 85741 Garching, Germany}

\begin{abstract}
$SU(2)$ gauge fields and axions can have a stable, isotropic and homogeneous configuration during inflation. However, couplings to other matter species lead to particle production, which in turn induces backreaction on and destabilization of the non-abelian and axion background. In this paper, we first study the particle production by a $SU(2)$ gauge field coupled to a massive Dirac doublet. To carry out this calculation we have made two technical improvements compared to what has been done in the literature. First, we apply the anti-symmetrization of the operators to treat particles and anti-particles on equal footing, second, to deal with the UV divergences, we apply instantaneous subtraction. We find that, the backreaction of produced fermions on the $SU(2)$ background is negligible for model parameters of observational interest. Next, we consider production of fermions due to coupling to the axion. The tree-level backreaction on the gauge fields, as well as on the axion, is vanishingly small. We also provide an estimate for the loop effects.
\end{abstract}

\pacs{Valid PACS appear here}
\maketitle


\section{Introduction}\label{sec:Intro}

Coupled axion and $SU(2)$ gauge fields can develop non-trivial vacuum expectation values (VEVs) \cite{Maleknejad:2011jw,Maleknejad:2011sq,Adshead:2012kp,Adshead:2013nka} during cosmic inflation \cite{Guth:1980zm,Sato:1980yn,Linde:1981mu,Albrecht:1982wi}. These inflationary models have a rich phenomenology that is not shared by canonical single scalar field inflation models (see \cite{Maleknejad:2012fw} for a review). As was first discovered by one of the authors (A.M.), when the conformal symmetry of Yang-Mills theory is broken by an effective $(F\tilde F)^2$ term in the Lagrangian, non-abelian gauge fields acquire an isotropic and homogeneous background VEV solution during inflation \cite{Maleknejad:2011sq, Maleknejad:2011jw}. Since then, several inflationary
models with the $SU(2)$ VEV have been introduced and studied which share the above features
\cite{Adshead:2012kp, Adshead:2013nka, Maleknejad:2016qjz,
Adshead:2016omu, Dimastrogiovanni:2016fuu, Adshead:2017hnc}. 

This background gauge field can provide a source for particle production during inflation.
 Despite being energetically subdominant, the axion-$SU(2)$ gauge field can produce potentially detectable signals during inflation. It can generate stochastic backgrounds of chiral gravitational waves \cite{Maleknejad:2016qjz,Dimastrogiovanni:2016fuu,Dimastrogiovanni:2012ew,Dimastrogiovanni:2012st, Adshead:2013qp}, tensor non-Gaussianity \cite{Agrawal:2017awz,Agrawal:2018mrg,Dimastrogiovanni:2018xnn} and the observed matter asymmetry in the Universe \cite{Maleknejad:2016dci,Adshead:2017znw,Caldwell:2017chz, Adshead:2018doq}.\footnote{For a study of the nonlinear impact of the spectator sector on the comoving curvature perturbation during inflation see  \cite{Papageorgiou:2018rfx,Papageorgiou:2019ecb}}

The upcoming LiteBIRD \cite{Matsumura:2014,Hazumi:2019lys} and CMB Stage-4 experiments are set to provide further constraints on the axion-gauge fields \cite{Abazajian:2016yjj,Thorne:2017jft,Shandera:2019ufi}. That is why it is important to check if models are viable phenomenologically. In particular, it is critical to see if couplings to other matter species can give rise to particle production and backreaction on the axion-gauge field background, thereby affecting the sourcing of observational signals. In \cite{Lozanov:2018kpk}, a charged scalar field was coupled to the $SU(2)$ gauge field, and production of pairs of charged particles in the non-trivial gauge field background (known as the Schwinger effect \cite{PhysRev.82.664}) was studied in de Sitter spacetime. It was found that the backreaction of the scalar particles on the $SU(2)$ background is negligible in the viable parameter regions of the simplest axion-$SU(2)$ models. In \cite{Maleknejad:2018nxz}, the backreaction of the extra spin-2 field in this setup was analytically studied for all the inflationary models involving the $SU(2)$ gauge field.

In this paper we continue our investigation of particle production by the axion-$SU(2)$ gauge field during inflation. This time, we study coupling of the $SU(2)$ gauge field to a pair of massive Dirac fermions, $i\bar{\Psi}\slashed{D}\Psi$. We also add an interaction between axion and the axial fermionic current, $J^{\mu5}\partial_{\mu}\varphi/\Lambda$, which is naturally expected in this type of models. We then calculate the backreaction of the fermions on the $SU(2)$ and axion background dynamics during inflation, following the framework we have established in \citep{Lozanov:2018kpk, Maleknejad:2018nxz}.

Fermionic particle production in de Sitter spacetime and its backreaction implications have been studied in the context of $U(1)$ theories. The case of a non-trivial abelian gauge field background without an axion was studied in \cite{Hayashinaka:2016qqn}; a slowly evolving axion background with no gauge field interactions was studied in \cite{Adshead:2015kza,Adshead:2018oaa}; and a combination of the two, assuming a massless fermion, was studied in \cite{Domcke:2018eki}. 

As for the $SU(2)$ gauge field background, the simplest fermionic non-abelian model was studied in the recent work \cite{Domcke:2018gfr}, where a massless doublet of Dirac fields is coupled covariantly to the $SU(2)$ gauge field. The main aim of \cite{Domcke:2018gfr} is to describe the fermionic particle production due to the quantum (loop) effects from the Adler-Bell-Jackiw (or chiral) anomaly. Our work not only extends their study to models with massive fermions and coupling to the axion background, but also provides the first detailed analysis of the allowed parameter space by cosmological and backreaction constraints.

We find important differences between the abelian and non-abelian models. Most notably, the leading order fermionic backreaction on the $SU(2)$ background is significantly smaller than in the fermionic abelian \cite{Hayashinaka:2016qqn} (as well as in the scalar abelian \cite{Frob:2014zka,Kobayashi:2014zza,Sharma:2017ivh,Kitamoto:2018htg,Shakeri:2019mnt}) cases. We find that the tree-level backreaction on the axion field backgound vanishes, unlike in $U(1)$ theories \cite{Adshead:2015kza,Adshead:2018oaa,Domcke:2018eki}. We thus conclude that, the inflationary scenarios involving an axion-$SU(2)$ gauge field spectator sector remain healthy and unaffected when couplings to gauged fermions are present.

The organization of the paper is the following. In Section \ref{sec:setup} we introduce our model. Section \ref{sec:PP} deals with the evolution and production of fermions in the time-dependent axion-$SU(2)$ background in a de Sitter universe. The results for the induced fermionic backreaction are presented in Section \ref{sec:BR}. Section \ref{sec:concl} is devoted to discussions and concluding remarks. Most of the technical details, including lengthy derivations, notations and conventions are delegated to the appendices.

\section{Fermions in axion-$SU(2)$ gauge field inflation}\label{sec:setup}

In this work, we study the fermion production by a slowly-evolving homogeneous and isotropic $SU(2)$ gauge field during inflation. The class of inflationary models involving such $SU(2)$ VEV has been first introduced in \cite{Maleknejad:2011jw, Maleknejad:2011sq}. Since then several different realizations of this class of models have been introduced and studied, e.g. \cite{Adshead:2012kp, Adshead:2013nka}. See section 2 of \cite{Maleknejad:2018nxz} and the references therein for a recent review on the models so far in the literature.\\

We assume slow-roll inflation with the background FLRW metric
\bea
ds^2= a^2(\tau) ( - d\tau^2 + \delta_{ij}dx^i dx^j),
\eea
where $\tau$ is the conformal time, and the scale factor, $a(\tau)$, is related to the Hubble parameter, $H$, as
\bea
a(\tau) \simeq - \frac{1}{H\tau} \an H\simeq const.
\eea

Besides, we have a homogeneous and isotropic $SU(2)$ gauge field background generated by one of the possible realizations of this class of models. In the temporal gauge (${\bf{A}}_0=0$), we have in \cite{Maleknejad:2011jw, Maleknejad:2011sq}
\bea\label{GF-SU2}
{\bf{A}}_{i} = A^a_{i} {\bf{T}}_a = a(\tau) \psi(\tau) \delta^a_i {\bf{T}}_a,
\eea
where $\psi(\tau)\simeq const.$ during slow-roll inflation and ${\bf{T}}_a$ are the generators of the $SU(2)$ group
\bea
[  {\bf{T}}_a , {\bf{T}}_b ] = i \epsilon^c_{~ab} {\bf{T}}_c.
\eea
Therefore, the gauge field has an almost constant energy density during inflation.

To avoid clutter, we suppress the spacetime indices of the Dirac matrices and spinors, unless otherwise stated. For example, $\gamma^0$ is a $4\times 4$ matrix, which can act on the $4$-dimensional column spinor $\Psi^{1}$ or can be acted upon by the $4$-dimensional row spinor $\bar{\Psi}^{1}\equiv\Psi^{1\dagger}\gamma^0$. We will have to deal with $8$, $4$ and $2$-component spinors which are acted upon by $8\times8$, $4\times4$ and $2\times2$ matrices, respectively to this end, we adopt the following notation. If the spinor (or the matrix) is $8$ (or $8\times8$) dimensional, then it has a tilde$(\sim)$ on top. The notation for the $4$-dimensional spinor and matrix remains unaltered, whereas the $2$-dimensional ones are written in boldface. Finally, $\rm{I}_n$ represents the $n\times n$ identity matrix and the gamma matrices are in the Dirac representation unless otherwise stated.

We consider a charged doublet of Dirac fermions
\Beq
\label{eq:Higgs}
\tilde{\Psi}=\begin{pmatrix}\Psi^{1}\\ \Psi^{2}\end{pmatrix}\,,
\Eeq
with the free theory
\Beq
\label{eq:FullAction-}
S_{\rm{fermion}}=\int\text{d}^4x\sqrt{-g}\left[i \bar{\tilde{\Psi}} {\sD} \tilde{\Psi} -m\bar{\tilde{\Psi}}\tilde{\Psi}\right]\,,
\Eeq
where $\bar{\tilde{\Psi}}= {\bf{(}} \bar{\Psi}_1 ~~ \bar{\Psi}_2 {\bf{)}} $, and $\slashed{D}$ is 
\Beq\label{slashD}
\slashed{D} \equiv {D}_{\mu} \otimes \gamma^{\mu} = {\bf{e}}^{\mu}_{~\alpha} \big[ \I \nabla_{\mu}  - i\ga A^a_{\mu} {\bf{T}}_a\big] \otimes \gamma^\alpha,
\Eeq
in which $\otimes$ is the Kronecker product, ${\bf{e}}^{\mu}_{~\alpha}= a^{-1}(\tau)\delta^{\mu}_{\alpha}$ are the vierbeins given by $g^{\mu\nu}={\bf{e}}^{\mu}_{~\alpha}{\bf{e}}^{\nu}_{~\beta}\eta^{\alpha\beta}$, and $\gamma^{\alpha}$ are the flat space Dirac matrices 
\bea
 \{\gamma^\alpha,\gamma^\beta\}=2\eta^{\alpha\beta}\II.
\eea
See Appendix \ref{Math} for details. Notice that $\gamma^\mu = {\bf{e}}^\mu_{~\alpha} \gamma^{\alpha} = a(\tau) \delta^{\mu}_{\alpha} \gamma^{\alpha}$ where $\alpha$ and $\beta$ represent the local Lorentz indices while $\mu$ and $\nu$ represent spacetime coordinates.

Since the inflationary setups in this paper involve axion fields, fermionic sector can have the following effective interaction with the axion (see for instance \cite{Weinberg:1996kr})
\Beq
\label{eq:Int-Action}
S_{\rm{int}}=\int\text{d}^4x\sqrt{-g}\left[\beta \frac{\lambda \varphi}{f}  \nabla_{\mu} J^{\mu}_5\right]\,,
\Eeq
where $\varphi$ is the axion field, $f$ is the axion decay constant, $\lambda$ is the dimensionless coefficient of the Chern-Simons interaction term of the axion, and $\beta$ is a dimensionless parameter. The quantity $\beta$ can be of order unity. Here, $J^{\mu}_5$ is the fermionic chiral current given as
\bea\label{Jmu5}
J^{\mu}_5 \equiv  \bar{\tilde{\Psi}} \I\otimes (\gamma^\mu \gamma^5)\tilde{\Psi},
\eea
where $\gamma^5=i\gamma^0\gamma^1\gamma^2\gamma^3$.

In summary, the fermion theory in an $SU(2)$-axion background given in \eqref{GF-SU2} and $\varphi = \varphi(\tau)$ is specified by

\Beq
\label{eq:calL}
S &=\int a^4\text{d}\tau \text{d}x^3  \frac{i}{a}  \bar{\tilde{\Psi}}\bigg[ \I \big( \p_{\tau} + \frac32 \mH \big)\otimes \gamma^0 \\ & +  \delta^{i}_{\alpha} \big( \I \p_i - i a \ga \psi \delta^a_i {\bf{T}}_a\big)\ \otimes \gamma^{\alpha} \\ & + i am \III 
+ \beta \frac{i \lambda \p_{\tau}\varphi}{f} \I \otimes (\gamma^0 \gamma^5) \bigg] \tilde{\Psi},
\Eeq
where the second term in the right hand side comes from the spin connection (see \eqref{Spin-dee} and \eqref{Spin-connect-FRW}). As implied by the above action, the canonically normalized field is
\bea
\tilde\Uppsi \equiv a^{\frac32}\tilde\Psi.
\eea
Using ${\bf{T}}_a=\frac12 \boldsymbol{\sigma}^i \delta^a_i$ ($\boldsymbol{\sigma}^i $ are the Pauli matrices) in \eqref{eq:calL}, we find
\bea
\label{theory}
\begin{split}
\mathcal{L} &=& i \bar{\tilde\Uppsi} \bigg[ \I  \p_{\tau} \otimes \gamma^0 +  \bigg( \I \p_i - \frac{i}{2}  \xa \mH \boldsymbol{\sigma}^i \bigg) \otimes \gamma^{i} \\ & &  + i  \mu_{{\rm m}} \mH {\III}  + 2i \xi_{\varphi} \mH \I \otimes (\gamma^0 \gamma^5) \bigg] \tilde\Uppsi,\,
\end{split}
\eea
where $\xa$, $\xi_{\varphi}$, and $\mu_{{\rm m}}$ are dimensionless parameters defined by
\bea
\xa \equiv \frac{\ga \psi}{H}, \quad \xi_{\varphi} \equiv \beta \frac{\lambda\p_{\tau}\varphi}{2afH}, \quad \mu_{{\rm m}} \equiv \frac{m}{H}.
\eea
For the sake of completeness, here we define another related dimensionless quantity in the axion inflation backgrounds 
\bea
\xi \equiv \frac{\lambda\p_{\tau}\varphi}{2afH},
\eea
which during slow-roll inflation is related to $\xp$ as $\xi \simeq \frac{1+\xp^2}{\xp}$ in the massless $SU(2)$ models. In our setup, $\xi$ and $\xpi$ are related as
\bea
\xpi = \beta \xi,
\eea
where $\beta$ is of order unity.

Up to this point, we wrote the theory in the flavor frame in terms of an 8-spinor in real space. However, in Fourier space, the setup is reducible into 2 irreducible 4-spinor sub-sectors in the helicity representation. Therefore, it is more convenient to go to Fourier space and write it in the extended helicity frame which we introduce now.\\
In Fourier space, we expand $\tilde\Uppsi$ as
\bea
\tilde\Uppsi(\tau,{\bf{x}}) = \int {\rm{d}}^3k e^{i{\bf{k}}.{\bf{x}}} \tilde\Uppsi_{{\bf{k}}}.
\eea
For a given momentum, ${\bf{k}}$, the $8\times 8$ helicity projection operators are
\Beq
\label{eq:P}
\tilde P_{\pm}({\bf{k}}) = \I \otimes \bigg(  \frac{ {\II} \pm k^i  \gamma^i }{2} \bigg),
\Eeq
which decompose $\tilde\Uppsi_{{\bf{k}}}$ into the plus and minus helicity states as
\bea
\tilde\Uppsi^{\pm}_{{\bf{k}}} = \tilde P_{\pm}({\bf{k}}) \tilde\Uppsi_{{\bf{k}}},
\eea
where $\tilde\Uppsi_{{\bf{k}}}=\tilde\Uppsi^{+}_{{\bf{k}}}+\tilde\Uppsi^{-}_{{\bf{k}}}$.
The helicity representation decomposes the system of 8-spinor in real space in Eq. \eqref{theory} into two subspaces of 4-spinors in real space
\bea
\tilde\Uppsi = \Uppsi^+ \oplus \Uppsi^- = \begin{pmatrix}
\Uppsi^+ \\
\Uppsi^-
\end{pmatrix},
\eea
 such that the theory is given as
$$S[\tilde\Uppsi] = S_+[\Uppsi^+] + S_-[\Uppsi^-].$$
We present the details of our consideration  in Appendix  \ref{helicity}. Here, we write the final theories for each of the 4-spinor subspaces in Fourier space \footnote{Note that the form of the $S_+$ is very similar to the action describing a {\it single} fermion derivatively coupled to an axion considered in  \cite{Adshead:2018oaa}. There the authors make a local chiral transformation of the fermion basis to avoid unphysical behaviour of the Bogoliubov coefficients in the massless fermion. However, for sufficiently large masses $(\mu_m \gtrsim 1)$ the transformation is not necessary. We do not follow this prescription for two reasons. First, the focus of the current study is limited to $(\mu_m \gtrsim 1)$. Second, the transformation is not applicable in the presence of our non-abelian gauge field. In particular, it does not simplify the $S_-$ action, because of the last term in the action.\label{footnote2}}

\Beq
\label{eq:SLSR-P}
S_+=\int \text{d}\tau\text{d}^3k \bar{\Uppsi}^+_{\bk} \Bigg[i\gamma^0 \partial_{\tau}-&\gamma^3k-\big(2\xpi-\frac{\xi_A}{2}\big)  \mathcal{H}\lambda_4 \\&- \mu_{{\rm m}}\mH \II \Bigg]\Uppsi^+_{\bk}\,,
\Eeq

\Beq
\label{eq:SLSR-M}
S_-=\int \text{d}\tau\text{d}^3k \bar{\Uppsi}^-_{\bk} &\Bigg[i\gamma^0\partial_{\tau}+\gamma^3k-\big(2\xpi +\frac{\xi_A}{2}\big)\mathcal{H}\lambda_4 \\
&-\mu_{{\rm m}} \mH \II + \gamma^1\xi_A\mathcal{H}\Bigg] \Uppsi^-_{\bk} \,,
\Eeq
where
\Beq
\gamma^0=\begin{pmatrix*}[c] \I &\boldsymbol{0} \\ \boldsymbol{0}& -\I\end{pmatrix*}\,,\quad \gamma^i=\begin{pmatrix*}[c] \boldsymbol{0} & \boldsymbol{\sigma}^i \\ -\boldsymbol{\sigma}^i & \boldsymbol{0}\end{pmatrix*} \,, \lambda_{4}=\begin{pmatrix*}[c] \boldsymbol{0} & \I \\ -\I & \boldsymbol{0}\end{pmatrix*}\,.
\Eeq
The field equations of $\Uppsi^{+}_{\bf{k}}$ and $\Uppsi^{-}_{\bf{k}}$ are
\Beq\label{EOM-psi-p}
\Big[i\gamma^0 \partial_{\tau}-\gamma^3k-\big(2\xpi -\frac{\xi_A}{2}\big)\mathcal{H}\lambda_4 - \mu_{{\rm m}}\mH \II \Big]\Uppsi^+_{\bk}=0,
\Eeq
and 
\Beq\label{EOM-psi-m}
\Big[i\gamma^0\partial_{\tau}+\gamma^3k-\big(&2\xpi +\frac{\xi_A}{2}\big)\mathcal{H}\lambda_4 \\
&-\mu_{{\rm m}} \mH \II +\gamma^1\xi_A\mathcal{H}\Big] \Uppsi^-_{\bk}=0,
\Eeq
respectively. Therefore, we have two independent Dirac fermions, $\Uppsi^{\pm}_{\bf{k}}$. We solve them in the next section. Before that, let us take a closer look at the field equations to have a qualitative understanding of each field.

Our Dirac fields can be expanded as
\bea
\Uppsi^{\pm}_{\bf{k}} = \sum_{s=\pm} \begin{pmatrix}
~\uppsi^{\pm\uparrow}_{s}(\tau,k) {\bf{E}}_s \\
\\
s \uppsi^{\pm\downarrow}_{s}(\tau,k)  {\bf{E}}_s
\end{pmatrix},
\eea
where $\uppsi^{\pm\uparrow}_{s}(\tau,k)$ and $\uppsi^{\pm\downarrow}_{s}(\tau,k)$ are mode functions and ${\bf{E}}_s$ with $s=\pm 1$ are the two-spinor polarization states
\bea\label{E+-E-}
{\bf{E}}_+ = \begin{pmatrix}
1 \\ 0
\end{pmatrix} \an {\bf{E}}_- = \begin{pmatrix}
0 \\ 1
\end{pmatrix}.
\eea
Since we are already in the helicity states of the given momentum $\bk$, the 2-spinor polarization states are $k$-independent.\\
Using the above in the field equations \eqref{EOM-psi-p} and \eqref{EOM-psi-m}, we find that:

\begin{itemize}
\item{For the plus spinor field:~the field is decoupled in terms of the polarization spinor ${\bf{E}}_{s}$. Thus, we have two pairs of coupled field equations for each polarization.}
\item{For the minus spinor field:~the field equation is not diagonalizable in terms of ${\bf{E}}_{s}$. That is because of the extra (time dependent) term proportional to $\gamma^1$ in the minus field equation \eqref{EOM-psi-m}. As a result, we have four coupled field equations. In the limit of either being well inside the horizon, i.e. $k\gg \mH$, or $\xa\ll 1$, this term is negligible and the system reduces to two pairs of coupled field equations.}
\end{itemize}

\section{Fermion production}\label{sec:PP}

We now calculate the evolution of the plus and minus fermionic fields, $\Uppsi^{\pm}_{\bk}$. Since these 4-spinors are decoupled (see Eqs. \eqref{EOM-psi-p} and \eqref{EOM-psi-m}), we consider them separately. 

\subsection{$\Uppsi^+$ spinors}

We begin with the $\Uppsi^+_{\bk}$ modes, described by the $S_+$ action given in Eq. \eqref{eq:SLSR-P}. Since the modes are $4$-dimensional objects, their first order in time linear equation of motion, Eq. \eqref{EOM-psi-p}, should yield four linearly independent 4-spinor solutions, i.e.,
\Beq
\label{eq:psipls}
\Uppsi^+_{\bk}=\sum_{s=\pm}\left[U_{s,{\bf{k}}}^+(\tau)a^+_{s,\bk}+V_{s,-{\bf{k}}}^+(\tau)b^{+\dagger}_{s,-\bk}\right]\,,
\Eeq
where creation and annihilation operators satisfy 
\Beq
\label{eq:Anticommab-}
\{a^{+}_{s,\bk},a^{+\dagger}_{s',\bk'}\}&=\delta_{ss'}\delta^{(3)}(\bk-\bk')\,, \\ \{b^{+}_{s,\bk},b^{+\dagger}_{s',\bk'}\}&=\delta_{ss'}\delta^{(3)}(\bk-\bk')\,.
\Eeq
We then decompose $U_{s,k}^+$ and $V_{s,k}^+$ as
\Beq
\label{eq:UVpsipls}
U_{s,{\bf{k}}}^+(\tau)=\frac{1}{\sqrt{2}}\begin{pmatrix}{\bf E}_su^{\uparrow}_s(k,\tau) \\ s{\bf E}_su^{\downarrow}_s(k,\tau)\end{pmatrix}\,,
\Eeq
and
\Beq
 V_{s,-{\bf{k}}}^+(\tau)=\frac{1}{\sqrt{2}}\begin{pmatrix}{\bf E}_sv^{\uparrow}_s(k,\tau) \\ s{\bf E}_sv^{\downarrow}_s(k,\tau)\end{pmatrix}\,,
\Eeq
where ${\bf{E}}_s$ are the 2-spinor polarization states defined in \eqref{E+-E-}.

Using the procedure in Appendix \ref{Hamiltonian} we can derive the initial conditions, 
\Beq
\label{eq:uvpl}
u^{\uparrow}_{s}(k,\tau)=v^{\uparrow*}_{s}(k,\tau)\quad {\rm and} \quad u^{\downarrow}_{s}(k,\tau)=-v^{\downarrow*}_{s}(k,\tau)\,.
\Eeq
Since $u^{\uparrow\downarrow}_s$ and $v^{\uparrow\downarrow}_s$ depend on each other, we can first solve the field equations of $u^{\uparrow\downarrow}_s$, and then, use \eqref{eq:uvpl} to find $v^{\uparrow\downarrow}_s$ and read $V^{+}_{s,-{\bf{k}}}(\tau)$ as
\bea
V^{+}_{s,-{\bf{k}}}(\tau) = \frac{1}{\sqrt{2}}\begin{pmatrix}{\bf E}_s u^{\uparrow*}_{s}(k,\tau) \\ -s{\bf E}_s u^{\downarrow*}_{s}(k,\tau)\end{pmatrix}.
\eea
Upon substiting the ansatz \eqref{eq:psipls} and \eqref{eq:UVpsipls} into the field equation \eqref{EOM-psi-p}, we arrive at 2 sets of coupled field equations for each polarization
\bea \label{u-ups-eq}
&(i\partial_{\tau}-\mu_{{\rm{m}}}\mathcal{H})u^{\uparrow}_s-\left[k+s\left(2\xpi-\frac{\xi_A}{2}\right)\mathcal{H}\right]u^{\downarrow}_s=0\,,\\ \label{u-downs-eq}
&(i\partial_{\tau}+\mu_{{\rm{m}}}\mathcal{H})u^{\downarrow}_s-\left[k+s\left(2\xpi-\frac{\xi_A}{2}\right)\mathcal{H}\right]u^{\uparrow}_s=0\,.
\eea

To find analytical solutions we make the following decomposition
\Beq
\label{eq:uupdowngen}
u^{\uparrow}_s=\frac{1}{\sqrt{2\x}}\left(Y_s+Z_s\right)\quad{\rm and}\quad u^{\downarrow}_s=\frac{1}{\sqrt{2\x}}\left(Y_s-Z_s\right)\,,
\Eeq
where $\x$ is the physical momentum rescaled by $H$, i.e. ,
\Beq
\x\equiv\frac{k}{aH}=-k\tau\,.
\Eeq

The coupled set of first order differential equations \eqref{u-ups-eq} and \eqref{u-downs-eq} can be decoupled into two second order differential equations for $Y_s$ and $Z_s$ as
\bea\label{Y-eq-+}
&&\partial_{\x}^2 Y_s+\left[1-\frac{2i\kappa^+_s}{\x}+\frac{1/4-\mu^{+2}}{\x^2}\right]Y_s=0\,,\\ \label{Z-eq-+}
&&\partial_{\x}^2 Z_s+\left[1-\frac{2i\tilde\kappa^+_s}{\x}+\frac{1/4-\mu^{+2}}{\x^2}\right]Z_s=0\,,
\eea
where $\kappa^+_s$, and $\tilde{\kappa}^+_s$ are
\bea
\kappa^+_s=\frac{1}{2}+is\left(2\xpi-\frac{\xi_A}{2}\right)\,,
\eea
and
\bea
\tilde\kappa^+_s=-\frac{1}{2}+is\left(2\xpi-\frac{\xi_A}{2}\right),
\eea
while $\mu^{+}$ is 
\bea
\mu^{+}= i \bigg[ \mu_{{\rm{m}}}^2 + \big(2\xpi-\frac{\xi_A}{2}\big)^2 \bigg]^{\frac12}\,.
\eea 
The general solutions for equations \eqref{Y-eq-+} and \eqref{Z-eq-+} are $W_{\kappa,\mu}(-2i\x)$ and $M_{\kappa,\mu}(-2i\x)$ Whittaker functions. Setting the Bunch-Davies vacuum as the initial condition for $u^{\uparrow}_s$ and $u^{\downarrow}_s$ 
\bea\label{u-up-down-BD}
\lim_{\tau \rightarrow -\infty} u^{\uparrow}_s(\tau,k) = \frac{1}{(2\pi)^{\frac32}}\frac{e^{-ik\tau}}{\sqrt{2k}}\,,
\eea
and
\bea
\lim_{\tau \rightarrow -\infty} u^{\downarrow}_s(\tau,k) = \frac{1}{(2\pi)^{\frac32}}\frac{e^{-ik\tau}}{\sqrt{2k}},
\eea
and using the asymptotic form of the $W$ and $M$ functions in \eqref{WM-asymp}, we find that $Y_s$ and $Z_s$ are given by
\bea\label{eq.Ys-Zs}
Y_s = b_{1s} W_{\kappa_s^+,\mu^+}(-2i\x)\,,
\eea
and
\bea
Z_s = b_{2s} W_{\tilde\kappa_s^+,\mu^+}(-2i\x).
\eea
 Therefore, in the asymptotic past limit, we have $Z_s = \frac{i}{2} (b_{2s}/b_{1s}) \x^{-1} Y_s \ll Y_s$. Combination of the Bunch-Davies vacuum condition in \eqref{u-up-down-BD} and the asymptotic form of the $W$ function in \eqref{WM-asymp} gives $b_{1s}=\frac{1}{(2\pi)^{\frac32}}e^{s(\xp/4-\xpi)\pi}$. Moreover, subtracting \eqref{u-downs-eq} from \eqref{u-ups-eq} and keeping the dominant terms in the asymptotic past limit, we find $b_{2s}=-i\mu_{_{\rm{m}}}b_{1s}$. 
Finally, we have 
\Beq
Y_s &= \frac{1}{(2\pi)^{\frac32}} e^{s(\xp/4-\xpi)\pi} W_{\kappa_s^+,\mu^+}(-2i\x),\\
Z_s &= - \frac{i\mu_{_{\rm{m}}}}{(2\pi)^{\frac32}}  e^{s(\xp/4-\xpi)\pi} W_{\tilde\kappa_s^+,\mu^+}(-2i\x).
\Eeq
Note that the amplitudes and the relative phases of the Bunch-Davies vacuum modes are such that the corresponding Hamiltonian is diagonalized. We present the detailed calculation of the Hamiltonian in Appendix \ref{Hamiltonian}. We note that the Hamiltonian diagonilzation leaves a residual freedom in choosing the initial conditions for the mode functions. The same applies to the computation in the next section. In both cases, we chose a set of initial conditions for which the Hamiltonian is diagonalized in the sub-horizon limit.

\subsection{$\Uppsi^-$ spinors}

Next, we calculate the $\Uppsi^-_{\bk}$ modes, described by the $S_-$ action given in Eq. \eqref{eq:SLSR-M}. Since the modes are $4$-dimensional objects, their first order in time linear equation of motion, Eq. \eqref{EOM-psi-m}, should yield four linearly independent 4-spinor solutions, i.e.,
\Beq
\label{eq:psimns}
\Uppsi^-_{\bk}=\sum_{s=\pm}\left[U_{s,{\bf{k}}}^-(\tau)a^-_{s,\bk}+V_{s,-{\bf{k}}}^-(\tau)b^{-\dagger}_{s,-\bk}\right]\,,
\Eeq
where creation and annihilation operators satisfy 
\Beq
\label{eq:Anticommab-}
\{a^{-}_{s,\bk},a^{-\dagger}_{s',\bk'}\}&=\delta_{ss'}\delta^{(3)}(\bk-\bk')\,, \\
\{b^{-}_{s,\bk},b^{-\dagger}_{s',\bk'}\}&=\delta_{ss'}\delta^{(3)}(\bk-\bk')\,.
\Eeq

The Lagrangian of the minus subspace is not diagonalizable in a time independent frame. Hence, the helicity eigenstates are only the eigenstates of the Lagrangian in the asymptotic past limit. Therefore, we adopt a more general trial vector solutions 
\Beq
\label{eq:UVpsimns}
U_{s,k}^-(\tau)&=\frac{1}{\sqrt{2}}\begin{pmatrix}{\bf E}_su^{\uparrow}_{s,+}(k,\tau) \\ s{\bf E}_su^{\downarrow}_{s,+}(k,\tau)\end{pmatrix}\\
&+\frac{1}{\sqrt{2}}\begin{pmatrix}{\bf E}_{-s}u^{\uparrow}_{s,-}(k,\tau) \\ -s{\bf E}_{-s}u^{\downarrow}_{s,-}(k,\tau)\end{pmatrix}\,,\\
V_{s,k}^-(\tau)&=\frac{1}{\sqrt{2}}\begin{pmatrix}{\bf E}_sv^{\uparrow}_{s,+}(k,\tau) \\ s{\bf E}_sv^{\downarrow}_{s,+}(k,\tau)\end{pmatrix}\\
&+\frac{1}{\sqrt{2}}\begin{pmatrix}{\bf E}_{-s}v^{\uparrow}_{s,-}(k,\tau) \\ -s{\bf E}_{-s}v^{\downarrow}_{s,-}(k,\tau)\end{pmatrix}\,,
\Eeq
where ${\bf{E}}_s$ are the 2-spinor polarization states defined in \eqref{E+-E-}. Since in the asymptotic past limit the system is diagonalized in this particular basis, we have 
\bea
& \lim_{\x \rightarrow \infty} u^{\uparrow}_{s,-}(k\tau) = \lim_{\x \rightarrow \infty} u^{\downarrow}_{s,-}(k\tau) = 0,\\
& \lim_{\x \rightarrow \infty} v^{\uparrow}_{s,-}(k\tau) = \lim_{\x \rightarrow \infty} v^{\downarrow}_{s,-}(k\tau) = 0.
\eea
Using the procedure in Appendix \ref{Hamiltonian}, one can derive the initial conditions, 
\Beq
\label{eq:uv-}
u^{\uparrow}_{s,+}(k,\tau)=v^{\uparrow*}_{s,+}(k,\tau)\,,  u^{\downarrow}_{s,+}(k,\tau)=-v^{\downarrow*}_{s,+}(k,\tau)\,. 
 \Eeq

 Since $u^{\uparrow\downarrow}_{s,p}$ and $v^{\uparrow\downarrow}_{s,p}$ depend on each other, we can first solve the field equations of $u^{\uparrow\downarrow}_{s,p}$, and then, use \eqref{eq:uv-} to read $v^{\uparrow\downarrow}_{s,p}$ from their corresponding $u^{\uparrow\downarrow}_{s,p}$.

After the substitution of the ansatz from Eq. \eqref{eq:UVpsimns} into the equation of motion \eqref{EOM-psi-m}, we arrive at
\Beq
&(i\partial_{\tau}-\mu_{\rm{m}}\mathcal{H})u^{\uparrow}_{s,p}+\left[k-sp\left(2\xpi+\frac{\xi_A}{2}\right)\mathcal{H}\right]u^{\downarrow}_{s,p}\\&\qquad\qquad\qquad\qquad\qquad\qquad -sp\xi_{A}\mathcal{H}u^{\downarrow}_{s,-p}=0\,,\\
&(i\partial_{\tau}+\mu_{\rm{m}}\mathcal{H})u^{\downarrow}_{s,p}+\left[k-sp\left(2\xpi+\frac{\xi_A}{2}\right)\mathcal{H}\right]u^{\uparrow}_{s,p}\\&\qquad\qquad\qquad\qquad\qquad\qquad+sp\xi_{A}\mathcal{H}u^{\uparrow}_{s,-p}=0\,.
\Eeq
Note also that for each given $s$ and $p$, we have a pair of coupled equations between $p=+$ and $p=-$ fields. 

Following our approach from the previous section, we make the decomposition
\Beq
\label{eq:uupdowngenm}
u^{\uparrow}_{s,p}=\frac{1}{\sqrt{2\tilde{\tau}}}\left(Y_{s,p}+Z_{s,p}\right)
\Eeq
and 
\Beq
u^{\downarrow}_{s,p}=\frac{1}{\sqrt{2\tilde{\tau}}}\left(Y_{s,p}-Z_{s,p}\right)\,,
\Eeq
yielding
\Beq
\label{eq:YZ1storderm}
&i\partial_{\x} Y_{s,p}-\left[1-\frac{sp(4\xpi+\xi_A)-i}{2\x}\right]Y_{s,p}\\
&\qquad\qquad+\frac{\mu_{\rm{m}}}{\x}Z_{s,p}-\frac{sp\xi_{A}}{\x}Z_{s,-p}=0\,,\\
&i\partial_{\x} Z_{s,p}+\left[1-\frac{sp(4\xpi+\xi_A)+i}{2\x}\right]Z_{s,p}\\&\qquad\qquad+\frac{\mu_{\rm{m}}}{\x}Y_{s,p}+\frac{sp\xi_{A}}{\x}Y_{s,-p}=0\,.
\Eeq

We can reduce the above coupled first order equations to pairs of coupled second order equations
\Beq
\partial_{\x}^2 Y_{s,p}+\Bigg[1&+\frac{i-2sp(\xpi+\xi_A/2)}{\x}\\ \qquad\qquad &+\frac{1/4+\mu_{\rm{m}}^2+(\xpi+\xi_A/2)^2+\xi_A^2}{\x^2}\Bigg]Y_{s,p}\\
&\qquad-\frac{2\xi_A(\xpi+\xi_A/2)}{\x^2}Z_{s,-p}=0\,,
\Eeq
\Beq
\partial_{\x}^2 Z_{s,-p}+\Bigg[1&-\frac{i-2sp(\xpi+\xi_A/2)}{\x}\\ \qquad\qquad &+\frac{1/4+\mu_{\rm{m}}^2+(\xpi+\xi_A/2)^2+\xi_A^2}{\x^2}\Bigg]Z_{s,-p}\\
 &\qquad-\frac{2\xi_A(\xpi+\xi_A/2)}{\x^2}Y_{s,p}=0\,.
\Eeq
Unlike before, the system cannot be solved analytically. Therefore, we solve them numerically. Furthermore, the amplitudes and the phases of the modes are adjusted to diagonalize the Hamiltonian. For a detailed derivation of the Hamiltonian see Appendix \ref{Hamiltonian}.

\section{Backreactions}\label{sec:BR}

The action \eqref{eq:FullAction-} has a Noether current associated to the $SU(2)$ isospin
\bea\label{Jmua-}
J^{\mu a} = \delta^{\mu}_{\alpha} ~ \frac{g_{\!_A}}{2a^4}\bar{\tilde{\Uppsi}}~\boldsymbol{\sigma}^a\otimes\gamma^{\alpha}\tilde{\Uppsi},
\eea
and the axial vector current, $J^{\mu}_5$, given in \eqref{Jmu5}.
Notice that $J^{\mu a}\equiv \frac{\delta S_{\rm{fermion}}}{\sqrt{-g}\delta A^{a}_{\mu}}$ satisfies $\nabla_{\mu}J^{\mu a}\boldsymbol{\sigma}^a=\boldsymbol{0}$. 
The Noether current and divergence of the chiral current induce backreactions on the background 
field equations of an axion-$SU(2)$ setup. See \cite{Maleknejad:2018nxz} for details about a uniform presentation of the axion-$SU(2)$ class of models and in particular its section 2 for the background equations. The Noether 4-current backreacts on the background equation of the gauge field as
\Beq
\label{eq:QEoM}
\partial_{\tau}^2(a\psi)&+2\mathcal{H}\partial_{\tau}(a\psi)+(\partial_{\tau}\mathcal{H}+\mathcal{H}^2)(a\psi)\\&+2a^3g_{\!_A}^2\psi^3 - \frac{g_{\!_A}\lambda}{f}a^2\partial_{\tau}\varphi \psi^2 = -a^2\mathcal{J}\,,
\Eeq
where the spatially averaged component of the matter $3$-current is
\Beq
\label{eq:CurlyJ}
\mathcal{J}=\frac{1}{3a}\delta_{b}^{j}\langle J_{j}^{b}\rangle\,.
\Eeq
Moreover, the axial current backreacts on the axion background equation as
\Beq
\partial_{\tau}^2\varphi+2\mathcal{H}\partial_{\tau}\varphi+a^2\partial_{\varphi}V + 3\frac{g_{\!_A}\lambda}{f}a\psi^2(\mathcal{H}\psi+\partial_{\tau}\psi) = a^2\mathcal{B}\,,
\Eeq
where the backreaction term is defined as
\Beq
\label{eq:BAxionBack}
\mathcal{B}= \beta \frac{\lambda}{2f}\nabla_{\mu}J^{\mu}_5 =  - im\beta \frac{\lambda}{a^3f}\bar{\tilde{\Uppsi}}\gamma_5 \tilde{\Uppsi}\,.
\Eeq
The last equality uses the field equations of the fermions.\\

We begin with an outline of our prescription for computing the VEVs of quadratic fermionic quantum operators such as $\mathcal{J}$ and $\mathcal{B}$. We then calculate the fermionic backreaction on the gauge field and the axion backgrounds. 

\subsection{VEVs of quadratic fermionic operators}

In order to compute VEVs of quantum operators we follow the \cite{parker2009quantum,Figueroa:2013vif}. \\ 

Consider a general four component fermionic field, similar to the $\Uppsi^+$ and $\Uppsi^-$ defined in section \ref{sec:setup}. 
\Beq
\label{eq:4compfield}
\eta_{\alpha}({\bx,\tau})&=\int {\rm d}^3k \,\eta_{\bk,\alpha}(\tau)e^{i{\bk}\cdot{\bx}}\,,\\
\eta_{\bk,\alpha}(\tau)&=\sum_{s=\pm}\left[U_{s,\bk,\alpha}(\tau)c_{s,\bk}+V_{s,-\bk,\alpha}(\tau)d^{\dagger}_{s,-\bk}\right]\,,
\Eeq
with a quadratic action
\Beq
S_{\eta}=\int d\tau {\rm d}^3k\,\left[i\eta_{\bk,\alpha}^{\dagger}\partial_{\tau}\eta_{\bk,\alpha}-\eta_{\bk,\alpha}^{\dagger}\Omega_{\alpha\beta}(k,\tau)\eta_{\bk,\beta}\right]\,,
\Eeq
and a Hamiltonian
\Beq
H_{\eta}=\int {\rm d}^3k\,\eta_{\bk,\alpha}^{\dagger}\Omega_{\alpha\beta}(k,\tau)\eta_{\bk,\beta}\,,
\Eeq
where $\alpha$ and $\beta$ run from $1$ to $4$, and $c_{s,\bk}$ and $d_{s,-\bk}$ are time-independent particle and anti-particle annihilation operators. Using Eq. \eqref{eq:4compfield} in the Hamiltonian, we obtain
\Beq
H_{\eta}=\int{\rm d}^3k({\bf c}_{\bk}^{\dagger},{\bf d}_{-\bk})\begin{pmatrix}{\bf \mathcal{E}}^U & {\bf \mathcal{F}}^{\dagger} \\ {\bf \mathcal{F}} & {\bf \mathcal{E}}^V\end{pmatrix}\begin{pmatrix}{\bf c}_{\bk}\\{\bf d}_{-\bk}^{\dagger}\end{pmatrix}\,,
\Eeq
where
\Beq
{\bf c}_{\bk} = [c_{+,\bk} \quad c_{-,\bk}]^T \,,\quad {\bf d}_{\bk} = [d_{+,\bk} \quad d_{-,\bk}]^T\,,
\Eeq
and
\Beq
{\bf \mathcal{E}}^U_{ss'}=U_{s,\bk,\alpha}^*\Omega_{\alpha\beta}U_{s',\bk,\beta}\,,\quad {\bf \mathcal{F}}_{ss'}=V_{s,\bk,\alpha}^*\Omega_{\alpha\beta}V_{s',\bk,\beta}\,.\\
\Eeq

In the ground state of a given $k$-mode of the fermionic field, i.e., in the Bunch-Davies vacuum, the state vector is $|0_{\rm {BD}}\rangle$ 
\Beq
{\bf c}_{\bk,s}|0_{\rm {BD}}\rangle=0\,,\quad{\bf d}_{\bk,s}|0_{\rm {BD}}\rangle=0\,,
\Eeq
and
\Beq
{\rm BD}\,{\rm vacuum:}\quad {\bf \mathcal{E}}^U_{ss'}=-{\bf \mathcal{E}}^V_{ss'}=\omega_s\delta_{ss'}\,, \quad {\bf \mathcal{F}}_{ss'}=0\,.
\Eeq
When the mode is excited, its ${\bf \mathcal{E}}^{U,V}_{ss'}$ can attain non-zero off-diagonal components and its ${\bf \mathcal{F}}_{ss'}$ can also become non vanishing. However, we can still diagonalize ${\bf \mathcal{E}}^{U,V}_{ss'}$ with vanishing ${\bf \mathcal{F}}_{ss'}$ by re-writing the Hamiltonian as 
\Beq
H_{\eta}=\int{\rm d}^3k(\check{{\bf c}}_{\bk}^{\dagger},\check{{\bf d}}_{-\bk})\begin{pmatrix}{\bf \check{\mathcal{E}}}^U & {\bf \check{\mathcal{F}}}^{\dagger} \\ {\bf \check{\mathcal{F}}} & {\bf \check{\mathcal{E}}}^V\end{pmatrix}\begin{pmatrix}\check{{\bf c}}_{\bk}\\\check{{\bf d}}_{-\bk}^{\dagger}\end{pmatrix}\,,
\Eeq
where
\Beq
\label{eq:BogGen}
\begin{pmatrix}\check{\bf{c}}_{\bk}^{-}(\tau)\\ \check{\bf{d}}_{-\bk}^{-\dagger}(\tau)\end{pmatrix}=P_k(\tau)\begin{pmatrix}{\bf c}_{\bk}^{-}\\{\bf d}_{-\bk}^{-\dagger}\end{pmatrix}\,,
\Eeq
and
\Beq
P_k(\tau)P_k^{\dagger}(\tau)={\rm I}_{4}\,.
\Eeq
We choose the eigenvectors of
\Beq
\begin{pmatrix}{\bf \mathcal{E}}^U & {\bf \mathcal{F}}^{\dagger} \\ {\bf \mathcal{F}} & {\bf \mathcal{E}}^V\end{pmatrix}\,,
\Eeq
as the columns of $P_k^{\dagger}(\tau)$. Then,
\Beq
{\bf \check{\mathcal{E}}}^U_{ss'}=-{\bf \check{\mathcal{E}}}^V_{ss'}=\check{\omega}_s\delta_{ss'}\,, \quad {\bf \check{\mathcal{F}}}_{ss'}=0\,,
\Eeq
as promised. \\ 

After the transformation in Eq. \eqref{eq:BogGen}, a new state vector has the properties of the physical vacuum, $|0_{\tau}\rangle$, defined as  
\Beq
\check{{\bf c}}_{\bk,s}|0_{\tau}\rangle=0\,,\quad\check{{\bf d}}_{\bk,s}|0_{\tau}\rangle=0\,.
\Eeq
Note that we work in the Heisenberg picture in which the state vector is constant, i.e., it remains $|0_{\rm {BD}}\rangle$ throughout. 

The fermionic field can be re-written as
\Beq
\label{eq:4compfieldtau}
\eta_{\bk,\alpha}(\tau)&=\sum_{s=\pm}\left[\check{U}_{s,\bk,\alpha}(\tau)\check{c}_{s,\bk}+\check{V}_{s,-\bk,\alpha}(\tau)\check{d}^{\dagger}_{s,-\bk}\right]\,,
\Eeq
where
\Beq
\check{U}_{s,\bk,\alpha}&=\sum_{s'=\pm}\left[U_{s',\bk,\alpha}P^{\dagger}_{\frac{3-s'}{2},\frac{3-s}{2}}+V_{s'-,\bk,\alpha}P^{\dagger}_{\frac{7-s'}{2},\frac{3-s}{2}}\right]\,,\\
\check{V}_{s,-\bk,\alpha}&=\sum_{s'=\pm}\left[V_{s',-\bk,\alpha}P^{\dagger}_{\frac{7-s'}{2},\frac{7-s}{2}}+U_{s',\bk,\alpha}P^{\dagger}_{\frac{3-s'}{2},\frac{7-s}{2}}\right]\,.\\
\Eeq

We wish to find the VEV of a Hermitian operator, which is quadratic in the fermionic field, of the general form
\Beq
\mathcal{O}&(\bx,\tau)=\\ 
&\int {\rm d}^3k\,{\rm d}^3k'\,e^{i(\bk-\bk')\cdot\bx}\eta^{\dagger}_{\bk,\alpha}(\tau)A_{\alpha,\beta}(\bk,\bk',\tau)\eta_{\bk',\beta}(\tau)\,,
\Eeq
with $A_{\alpha,\beta}^{\dagger}(\bk,\bk',\tau)=A_{\alpha,\beta}(\bk',\bk,\tau)$ by virtue of the hermitian nature of the $\mathcal{O}$ operator.\\

When computing the VEV, we need to make sure that
\begin{itemize}
\item[1.] Particles and anti-particles are treated on equal footing.
\item[2.] Only physical field excitations, i.e., those on top of the physical vacuum, contribute to the VEV.
\end{itemize}
To address the first point, we follow \cite{parker2009quantum} and define the anti-symmetrizied operator
\Beq
\label{eq:Anti}
\mathcal{O}_a&(\bx,\tau)\equiv\\
&\int {\rm d}^3k\,{\rm d}^3k'\,e^{i(\bk-\bk')\cdot\bx}A_{\alpha,\beta}(\bk,\bk',\tau)[\eta^{\dagger}_{\bk,\alpha}(\tau),\eta_{\bk',\beta}(\tau)]\\
&=\int {\rm d}^3k\,{\rm d}^3k'\,e^{i(\bk-\bk')\cdot\bx}A_{\alpha,\beta}(\bk,\bk',\tau)\\
&\qquad\times\frac{1}{2}\Big(\eta^{\dagger}_{\bk,\alpha}(\tau)\eta_{\bk',\beta}(\tau)-\eta_{\bk',\beta}(\tau)\eta^{\dagger}_{\bk,\alpha}(\tau)\Big)\,,
\Eeq
which has the same classical counterpart as $\mathcal{O}$. The difference is that when we take the BD VEV, $\mathcal{O}$ receives only contributions from terms with $d_{s,-\bk}d_{s,-\bk}^{\dagger}$, i.e., the anti-particle creation and annihilation operators, whereas 
$\mathcal{O}_a$ receives contributions from both $c_{s,\bk}c_{s,\bk}^{\dagger}$ and $d_{s,-\bk}d_{s,-\bk}^{\dagger}$.

To account for the second point, we follow \cite{Figueroa:2013vif} and we subtract from the BD VEV the expectation value with respect to $|0_{\tau}\rangle$. This way only non-vacuum field fluctuations contribute to the physical vacuum expectation value. 

To sum up, the VEV of $\mathcal{O}$ is defined as
\Beq
\langle\mathcal{O}(\bx,\tau)\rangle\equiv\langle0_{\rm BD}|\mathcal{O}_a(\bx,\tau)|0_{\rm BD}\rangle-\langle0_{\tau}|\mathcal{O}_a(\bx,\tau)|0_{\tau}\rangle\,,
\Eeq
which reduces to
\Beq
\label{eq:Oreg}
\langle\mathcal{O}&(\bx,\tau)\rangle=\int {\rm d}^3k\,A_{\alpha\beta}(\bk,\bk,\tau)\sum_{s=\pm}\frac{1}{2}\\
&\times\Big[\Big(V_{s,\bk,\alpha}^*(\tau)V_{s,\bk,\beta}(\tau)-U_{s,\bk,\beta}(\tau)U_{s,\bk,\alpha}^*(\tau)\Big)\\
& -\Big(\check{V}_{s,\bk,\alpha}^*(\tau)\check{V}_{s,\bk,\beta}(\tau)-\check{U}_{s,\bk,\beta}(\tau)\check{U}_{s,\bk,\alpha}^*(\tau)\Big)\Big]\,.
\Eeq

\begin{figure*}[th]
\includegraphics[width=3.0in]{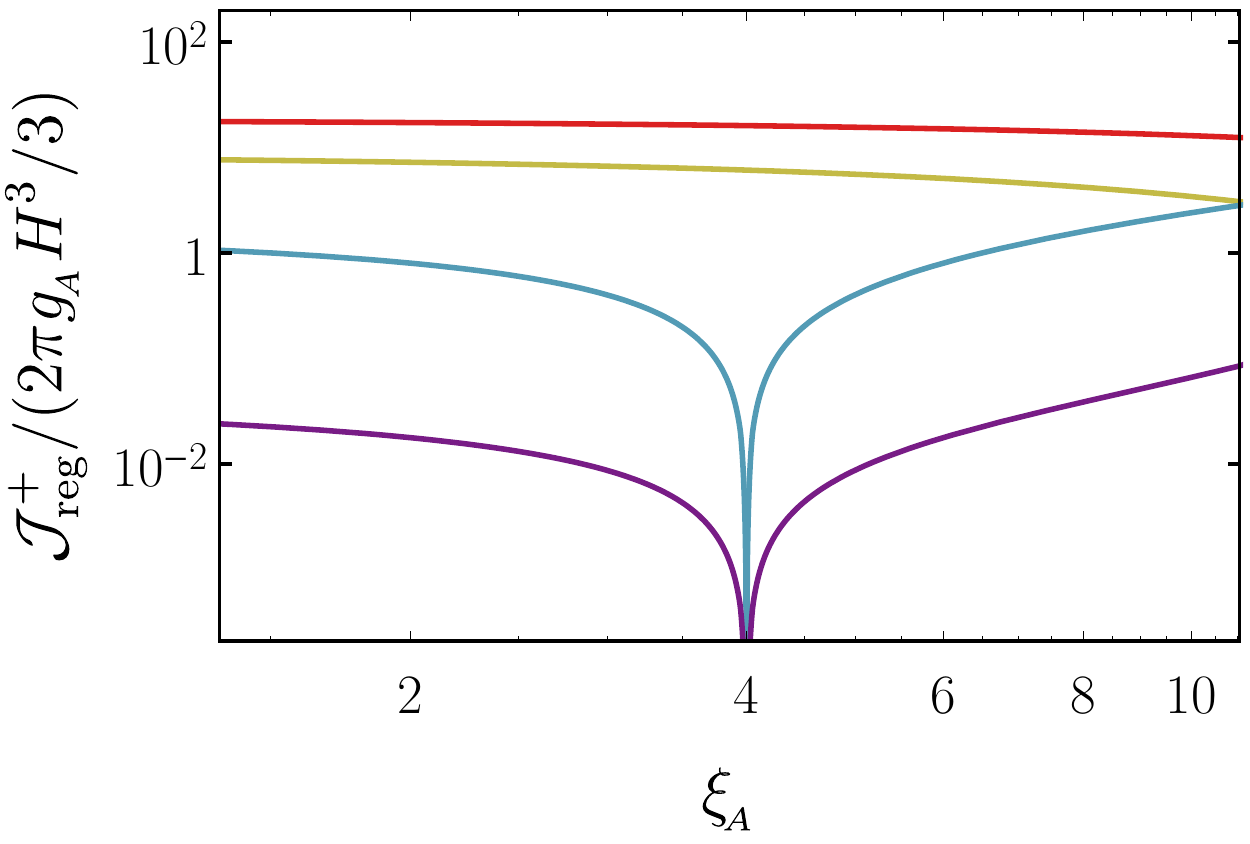}\hspace{0.5in}
\includegraphics[width=3.0in]{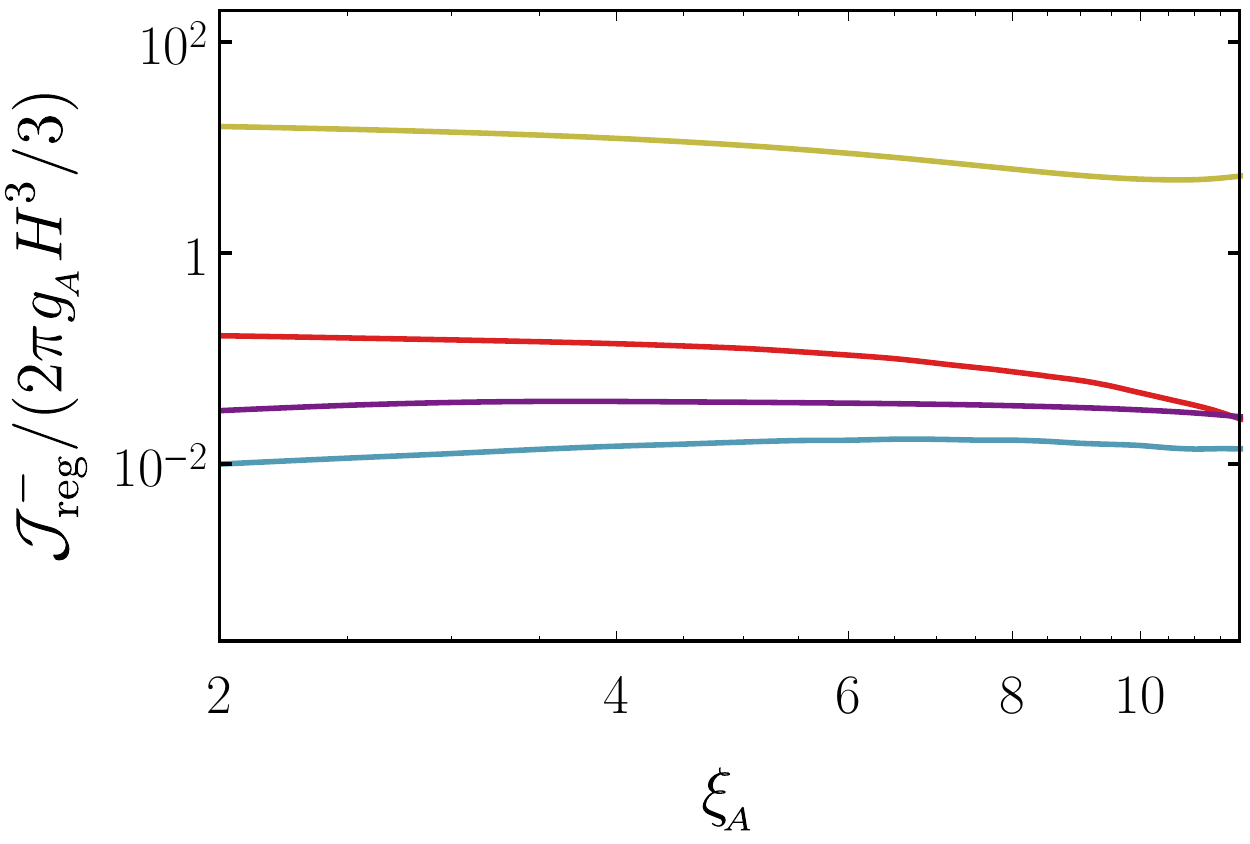}
\caption{The expectation values of the backreaction currents of the $+$ (top panel) and $-$ (bottom panel) fermions as a function of $\xi_{\!A}$ for $m=10H$, $\xi_{\varphi}=1$ (purple); $m=H$, $\xi_{\varphi}=1$ (blue); $m=H$, $\xi_{\varphi}=10$ (gold); $m=10H$, $\xi_{\varphi}=10$ (red). The prominent dip in the first panel at $2\xi_{
\varphi}-\xi_{\!A}/2=0$ is due to an exact cancelation between the effective masses induced by the gauge field and the axion. Such a cancelation is not observed in the $-$ fermions.}
\label{fig:JxiA}
\end{figure*}

Here, we are interested in the backreaction of the fermion on the gauge field field equation, $\mathcal{J}^{\pm}$, and the axion field equation, $\mathcal{B}^{\pm}$, which can be computed by the above  formula as 
\Beq
\label{calJ1}
\langle \mathcal{J}^s&\rangle=\int {\rm d}^3k\,A^{\mathcal{J}^s}_{\alpha\beta}(\bk,\bk,\tau)\frac{1}{2}\\
&\times\Big[\Big(V_{s,\bk,\alpha}^*(\tau)V_{s,\bk,\beta}(\tau)-U_{s,\bk,\beta}(\tau)U_{s,\bk,\alpha}^*(\tau)\Big)\\
& -\Big(\check{V}_{s,\bk,\alpha}^*(\tau)\check{V}_{s,\bk,\beta}(\tau)-\check{U}_{s,\bk,\beta}(\tau)\check{U}_{s,\bk,\alpha}^*(\tau)\Big)\Big]\,,
\Eeq
and 
\Beq
\label{calB-}
\langle \mathcal{B}^s&\rangle=\int {\rm d}^3k\,A^{\mathcal{B}^s}_{\alpha\beta}(\bk,\bk,\tau)\frac{1}{2}\\
&\times\Big[\Big(V_{s,\bk,\alpha}^*(\tau)V_{s,\bk,\beta}(\tau)-U_{s,\bk,\beta}(\tau)U_{s,\bk,\alpha}^*(\tau)\Big)\\
& -\Big(\check{V}_{s,\bk,\alpha}^*(\tau)\check{V}_{s,\bk,\beta}(\tau)-\check{U}_{s,\bk,\beta}(\tau)\check{U}_{s,\bk,\alpha}^*(\tau)\Big)\Big]\,,
\Eeq
respectively.\\

\subsection{Backreaction on the $SU(2)$ background}

We now calculate the homogeneous and isotropic backreaction term on the $SU(2)$ background, $\mathcal{J}$, defined in Eq. \eqref{eq:CurlyJ}. It conveniently separates into two independent contributions from the $+$ and $-$ fermions:
\Beq
\mathcal{J}=\mathcal{J}^++\mathcal{J}^-\,.
\Eeq
The expressions for $\mathcal{J}^{+}$ and $\mathcal{J}^{-}$ take the form given in Eq. \eqref{calJ1}, with
\Beq
A^{\mathcal{J}+}(\bk,\bk,\tau)&= ~~\frac{g_{\!A}}{3a^3}\begin{pmatrix*}[r]~~0 & ~~0 & ~~1 & ~~0 \\ ~~0 & ~~0 & ~~0 & ~~1 \\ ~~1 & ~~0 & ~~0 & ~~0 \\ ~0 & ~1 & ~0 & ~0 \end{pmatrix*}\,,\\
A^{\mathcal{J}-}(\bk,\bk,\tau)&=-\frac{g_{\!A}}{3a^3}\begin{pmatrix*}[r]0 & 0 & 1 & -2 \\ 0 & 0 & -2 & 1 \\ 1 & -2 & 0 & 0 \\ -2 & 1 & 0 & 0 \end{pmatrix*}\,.
\Eeq

In Fig. \ref{fig:JxiA}, we show $\mathcal{J}^{+}$(top) and $\mathcal{J}^{-}$(bottom) for different values of the parameters $\xi_{\varphi}$, $m$ and $\xi_{A}$. We observe the following dependence:

\begin{itemize}
\item $\mathcal{J}^{+}$ has a prominent dip when $2\xi_{\varphi}-\xi_{A}/2=0$, which occurs because the axion and gauge field-induced effective mass terms cancel. Besides this feature, for a fixed mass, $\mathcal{J}^{+}$ increases monotonically  with $\xi_A$. Otherwise when the "bare" mass of the fermion is the dominant scale, i.e., $m/H>\xi_{A}, \xi_{\varphi}$, we observe a decrease in particle production as the mass increases, as expected. In the opposite limit, $m/H<\xi_{A},\xi_{\varphi}$, there is an increase in particle production as the mass increases, until the mass becomes the dominant scale. \footnote{A similar increase in particle production with the increase in mass was observed in \cite{Adshead:2018oaa}. In their setup, the fermion is not coupled to a gauge field but is derivatively coupled to an axion field.}
\item $\mathcal{J}^{-}$ exhibits a complex behaviour with the parameters which we attribute to the additional couplings in this sector. Using the current regularization scheme and for the parameter region of interest, $\mathcal{J}^{-}$ never exceeds $\mathcal{J}^{+}$ apart from the dips in $\mathcal{J}^{+}$, so the dominant contribution to the backreaction considered in the next section comes from $\mathcal{J}^+$.
\item When compared to the scalar case considered in \cite{Lozanov:2018kpk}, the fermion model has an important new feature. Unlike scalars, fermion particles are copiously produced as $\xi_{A}$ increases and dominate the other scales in the problem. Our setup provides a novel mechanism for efficient production of fermionic matter during inflation.
\item The fermion Schwinger particle production by a homogeneous $U(1)$ gauge field studied in \cite{Hayashinaka:2016qqn} is different from our $SU(2)$ case with the isotropic and homogeneous VEV. In the $U(1)$ case the current decreases with the increase of the fermion mass.  However, in both cases, the current increases like $\xp^2$ in the very strong gauge field limit.
\end{itemize}

\begin{figure*}[t] 
 
  \includegraphics[height=4.2in]{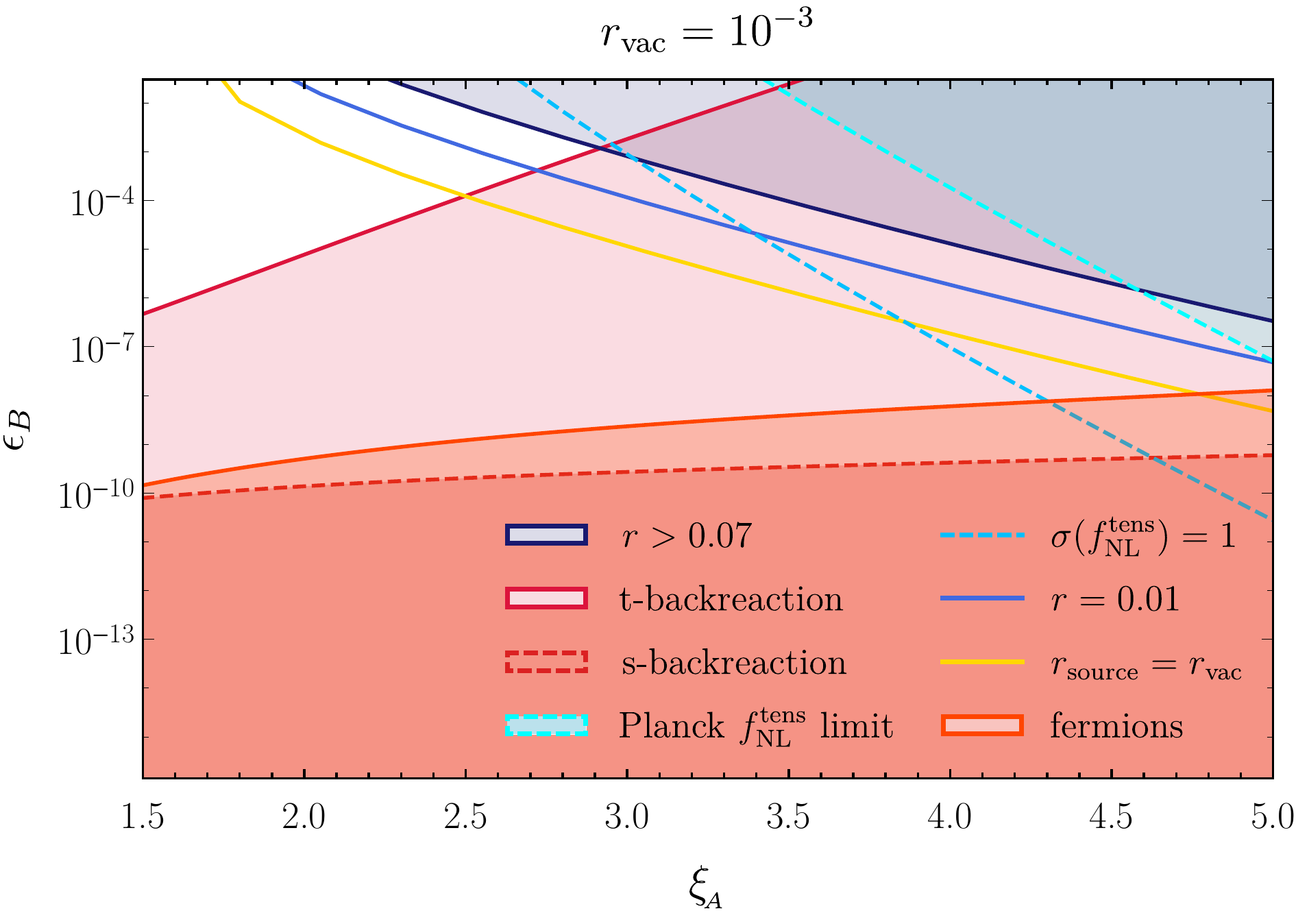} 
   \caption{We plot the energy density fraction of the gauge field $\epsilon_B$ as a function of the effective mass of the gauge field, $\xi_A$. Excluded parameter space with $r_{vac} = 10^{-3}$. The blue shaded area is excluded by the tensor-to-scalar ratio, the light red area by the large tensor backreaction discussed in \cite{Maleknejad:2018nxz}. The cyan area by the tensor non-Gaussianity, and the dark red area by the Schwinger pair-creation of scalar fields discussed in \cite{Lozanov:2018kpk}. The blue and yellow lines show $r = 10^{-2}$ and $r_{source} = r_{vac}$, respectively, while the dashed cyan line shows $f^{tens}_{NL} = 1$. The bound from the fermion particle production, depicted by the orange solid line (for $m = H$ and  $\xi_{\varphi}= 1$) and the area underneath, does not lead to additional bounds on the observationally relevant parameter space. The $\xi_A$ on the horizontal axis is the same as $m_Q$ in \cite{Dimastrogiovanni:2016fuu,Agrawal:2017awz, Agrawal:2018mrg}.}
   \label{fig:EpsBrvace-3xi1M1}   
\end{figure*}

\begin{figure*}[t] 

  \includegraphics[height=4.2in]{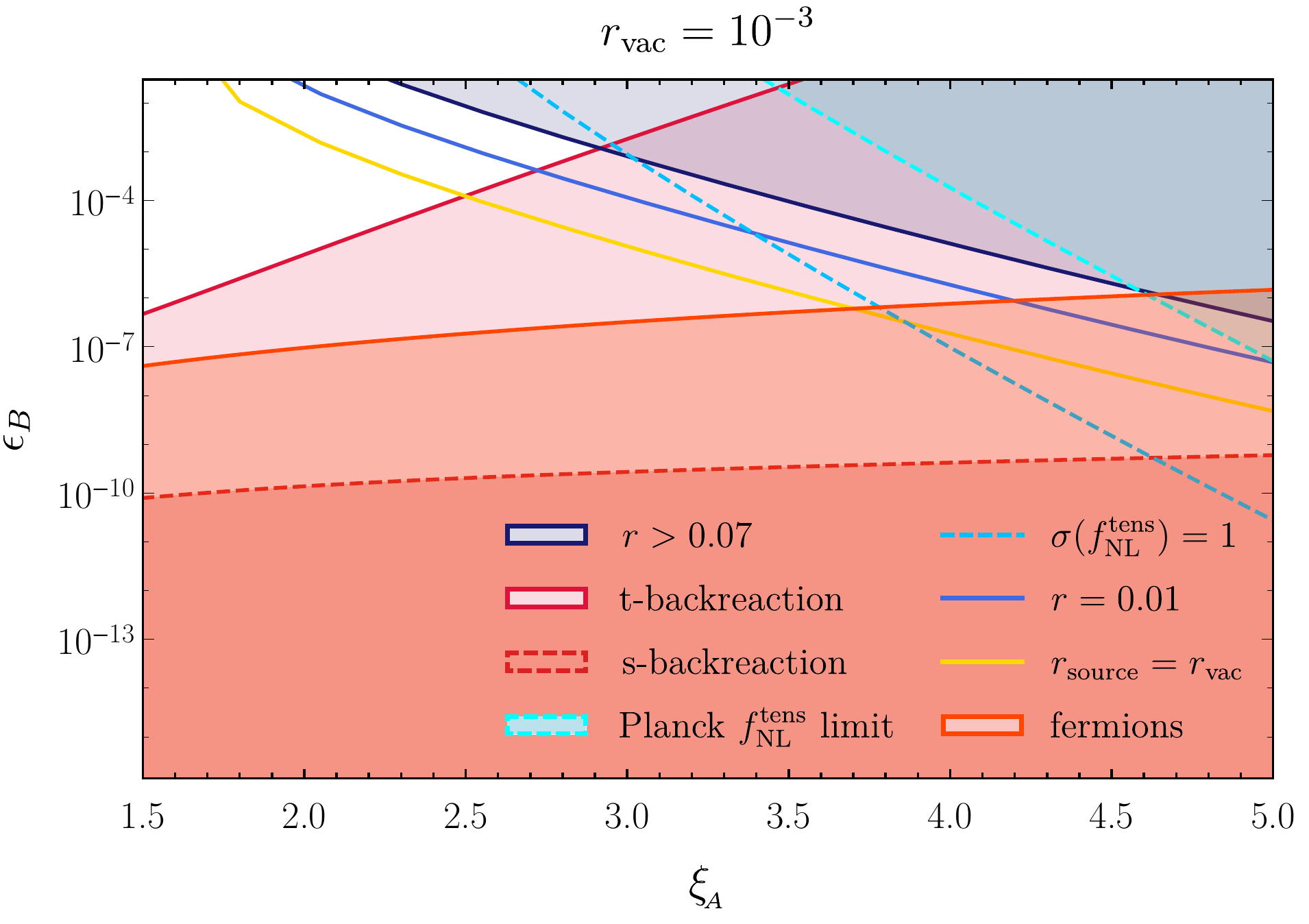} 
   \caption{Same as Fig. \ref{fig:EpsBrvace-3xi1M1}, but for $m=10H$ and $\xi_{\varphi}=10$.}
   \label{fig:EpsBrvace-3xi10M10}   
\end{figure*}

\begin{figure*}[t] 

  \includegraphics[height=4.2in]{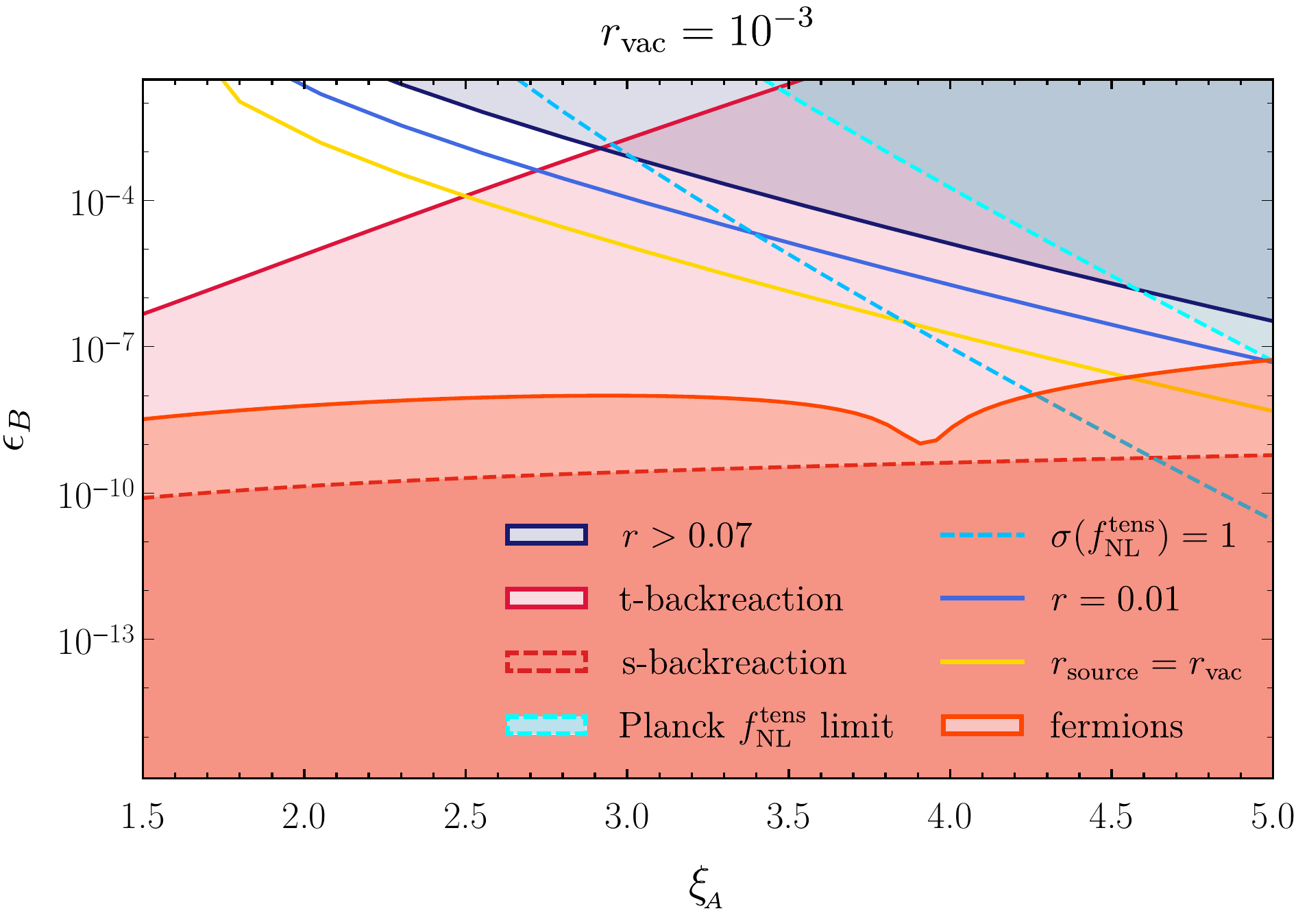} 
   \caption{Same as Fig. \ref{fig:EpsBrvace-3xi1M1}, but for $m=10H$ and $\xi_{\varphi}=1$.}
   \label{fig:EpsBrvace-3xi1M10}   
\end{figure*}

\begin{figure*}[t] 

  \includegraphics[height=4.2in]{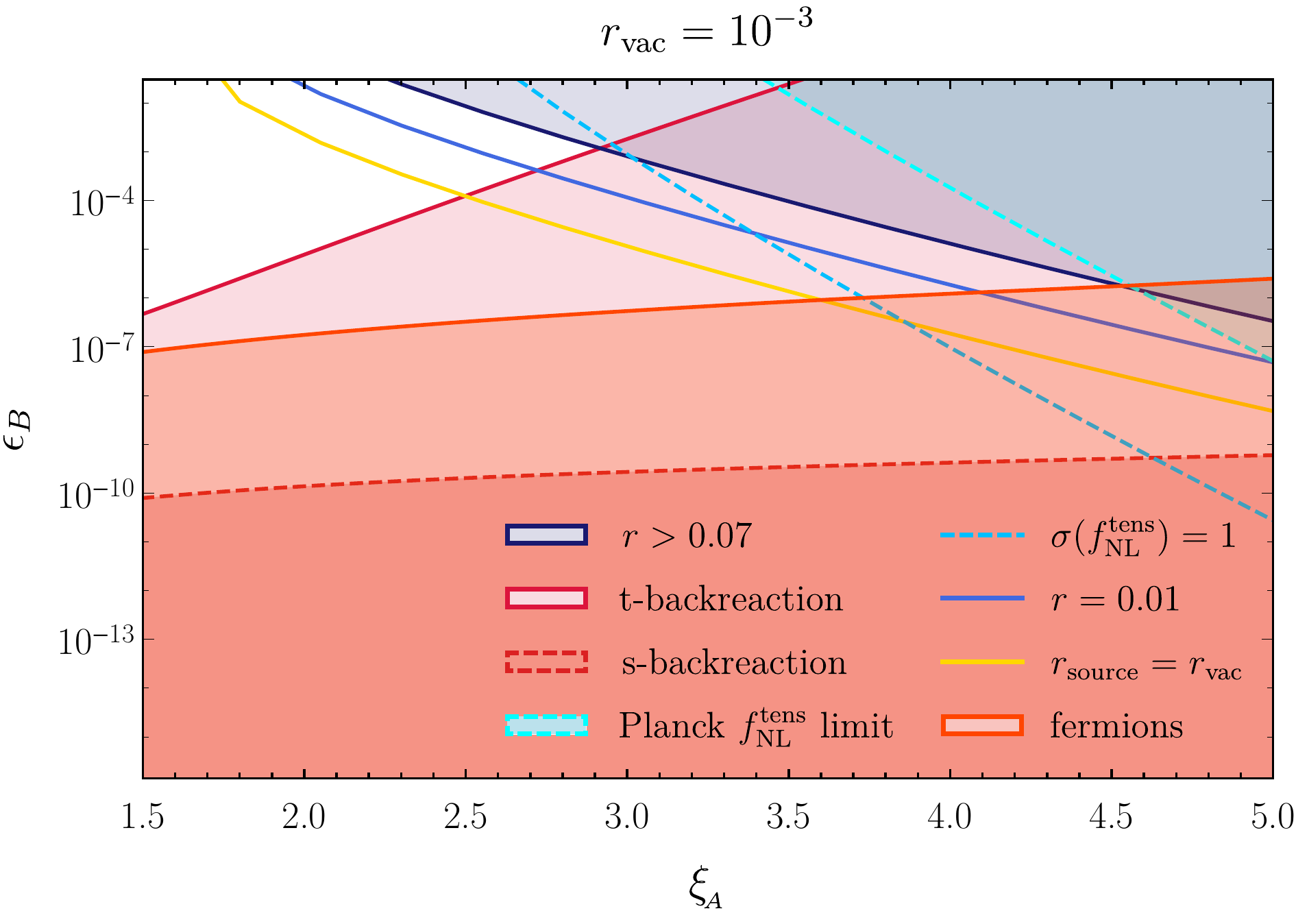} 
   \caption{Same as Fig. \ref{fig:EpsBrvace-3xi1M1}, but for $m=H$ and $\xi_{\varphi}=10$.}
   \label{fig:EpsBrvace-3xi10M1}   
\end{figure*}

Having computed $\mathcal{J}^{\pm}$, we can compute their backreaction on the $SU(2)$ gauge field background, $\psi$, by following \cite{Maleknejad:2018nxz}. Assuming slow-roll evolution of the axion-$SU(2)$, i.e., $\frac{\ddot\psi}{H^2\psi}\ll \frac{\dot\psi}{H\psi}\ll 1$, the field equation of the gauge field given in \eqref{eq:QEoM} can be approximated as
\bea\label{eq--psiIV}
3H \dot\psi + \dot{H}\psi + V_{{\rm{eff}},\psi}(\psi)\simeq 0,
\eea
where a dot is a derivative with respect to cosmic time and the field derivative of the effective potential of $\psi$ is 
\bea\label{eq--psiV}
V_{{\rm{eff}},\psi}(\psi) \simeq 2 H^2\psi(1+\xp^2) -  \frac{\ga\lambda\dot\varphi}{f}\psi^2.
\eea
Slow-roll demands $V_{{\rm{eff}},\psi}(\psi) \ll 1$, while
each of the terms in the right hand side can be much larger,
e.g., $\frac{\ga\lambda\dot\varphi}{f}\psi^2/V_{{\rm{eff}},\psi}\gg1
$. On the other hand, $\mathcal{J}$ in the right hand side of \eqref{eq:QEoM} should be at most on the order of the slow-roll suppressed terms, i.e., 
\bea\label{eq:JH}
\frac{\mathcal{J}}{H^2\psi}  \ll 1\,,
\eea
so that it does not break the slow-roll dynamics in the background and can be considered as a perturbation correction. Otherwise, the perturbative expansion and slow-roll dynamics are not trustable and the models should be studied numerically. 

Therefore, we define the regime of strong backreaction as
\Beq\label{eq:JH10}
\mathcal{J} < 10^{-2}H^2\psi\,.
\Eeq
We will use the above to explore the possible parameter space of one possible axion-$SU(2)$ gauge field model in section \ref{toy}.

\subsection{Backreaction on the axion background}

We now turn to the homogeneous and isotropic backreaction term on the axion background, $\mathcal{B}$, defined in Eq. \eqref{eq:BAxionBack}. It again splits into the sum of two independent $+$ and $-$ components:
\Beq
\mathcal{B}=\mathcal{B}^++\mathcal{B}^-\,.
\Eeq
$\mathcal{B}^{+}$ and $\mathcal{B}^{-}$ reduce to the form given in Eq. \eqref{calB-}
\Beq
A^{\mathcal{B}+}(\bk,\bk,\tau)=A^{\mathcal{B}-}(\bk,\bk,\tau)=\beta \frac{ \lambda mH^3}{2if}\begin{pmatrix*}[c]\boldsymbol{0} & - \I \\ \I & \boldsymbol{0} \end{pmatrix*}\,.
\Eeq

From the axion field equation in the slow-roll regime, we have $\frac{\ddot\varphi}{H\dot\varphi}\ll1$, and 
\bea
3H\dot\varphi + V_{\varphi,\rm{eff}} \simeq 0,
\eea
where $ V_{\varphi,\rm{eff}} = V_{\varphi} +\frac{3\lambda \ga}{f}\psi^2(\dot\psi+H\psi)$. 
The validity of the perturbation and slow-roll dynamics requires $\mathcal{B}$ to be at most of the order of the slow-roll suppressed terms, e.g. 
\bea\label{calB-c}
\mathcal{B} \ll H\dot\varphi.
\eea
 Interestingly, however we find that our choice of initial conditions that makes the Hamiltonian diagonalized yields 
\Beq
\label{calB--}
A^{\mathcal{B}^s}_{\alpha\beta}(\bk,\bk,\tau)\Big(V_{s,\bk,\alpha}^*(\tau)V_{s,\bk,\beta}(\tau)-U_{s,\bk,\beta}(\tau)&U_{s,\bk,\alpha}^*(\tau)\Big)\\
 &=0\,,
\Eeq
which implies that 
\bea
\mathcal{B} = 0.
\eea
Therefore, the particle production does not lead to {\it any} backreaction on the axion background. Note that this is the direct result of our quantization skim based on diagonalization of the Hamiltonian. Setting the
initial condition based on charge conjugation symmetry which does not diagonalize the Hamiltonian at asymptotic past, leads to a different and non-vanishing value for $\nabla_{\mu} J^{\mu}_5$, which is worked out in \cite{Maleknejad:2019hdr}. 

The computation so far has been done effectively at tree level. A one-loop effect, which has not been included consistently (see \cite{Domcke:2018gfr} for a related work with massless fermions), is the chiral anomaly, i.e., a quantum correction to the expectation value of $\nabla_{\mu}J^{\mu5}$, equal to $g_{A}^2{\bf{Tr}}(\boldsymbol{F}_{\mu\nu}\tilde{\boldsymbol{F}}^{\mu\nu})/(16\pi^2)$ \cite{Weinberg:1996kr} which is $\approx 3g_{A}^3\psi^3H/(4\pi^2)$. Since $\mathcal{B}=(\beta\lambda)/(2f)\nabla_{\mu}J^{\mu5}$, using $g_{\!A}\lambda\partial_{\tau}\varphi/(af)\approx2H(\xp+\xp^{-1})$, Eq. \eqref{calB-c}  yields

\Beq
\Big(\frac{f}{H}\Big)^2 \gg \frac{3}{16\pi^2}\beta\xi _A^2 (\xi _A^2+1) \,.
\Eeq

Since the right-hand side is always of order of unity, the backreaction is never important when $f \gg H$.

\subsection{Parameter space of a model}\label{toy}

Our method applies to models in which inflation is driven by the axion-gauge field sector  \cite{Maleknejad:2011jw,Maleknejad:2011sq,Adshead:2012kp,Adshead:2013nka}, as well as to those in which the axion and gauge fields are in a spectator sector \cite{Dimastrogiovanni:2016fuu}. For concreteness, we consider the latter model and compare our results with observational bounds on the following spectator model:
\Beq
\label{eq:Action}
&S=S_{\rm{EH}}+S_{\phi}+S_{\rm{spec}}+S_{\rm{fermion}}+S_{\rm{int}}\,,\\
&S_{\rm{spec}}=\int\text{d}^4x\sqrt{-g}\Bigg[\frac{1}{2}\partial_{\mu}\varphi\partial^{\mu}\varphi-V(\varphi)\\
&\qquad-\frac{1}{2}{\bf{Tr}(\boldsymbol{F}_{\mu\nu}\boldsymbol{F}^{\mu\nu})}-\frac{\lambda \varphi}{2f}{\bf{Tr}(\boldsymbol{F}_{\mu\nu}\tilde{\boldsymbol{F}}^{\mu\nu})}\Bigg]\,,
\Eeq
where $S_{\rm{EH}}$ and $S_{\phi}$ are the Einstein-Hilbert and the inflaton actions, respectively, responsible for inflation of the universe, and  $S_{\rm{spec}}$ is the action of the spectator sector. It contains the axion-gauge field sector, where $\varphi$ is the axion with a potential $V(\varphi)$ and a decay constant $f$, and 
\Beq
{\bf{F_{\mu\nu}}}=\nabla_{\mu}{\bf{A}_{\nu}}-\nabla_{\nu}{\bf{A}_{\mu}}+ig_{\!A}\left({\bf{A}_{\mu}}{\bf{A}_{\nu}}-{\bf{A}_{\nu}}{\bf{A}_{\mu}}\right)\,,
\Eeq
is the field strength tensor of the $SU(2)$ gauge fields. The last term in $S_{\rm{spec}}$ is the Chern-Simons interaction, where $\lambda$ parametrizes its strength and $\tilde{\bf{F}}^{\mu\nu}$ is the dual of $\bf{F}_{\mu\nu}$.

For the bound on the backreaction of the gauge fields we use Eq. \eqref{eq:JH10}, which reduces to
\Beq
\label{eq:SchwEpsB}
\epsilon_B < \xi _A ^3  \frac{10^2\pi^2A_{\rm{s}}r_{\rm{vac}}}{2}\frac{\mathcal{J}}{g_{\!A}H^3}\,,
\Eeq
where $\epsilon_B\equiv \xp^4H^2/(g_{\!A}^2m_{\rm{pl}}^2)$ is about two times the energy density fraction of the gauge field. We have also used the slow-roll relation $r_{\rm{vac}}=2H^2/(A_{\rm{s}}\pi^2m_{\rm{pl}}^2)$ to parametrize the Hubble scale of inflation, i.e., $r_{\rm vac}$ is the standard vacuum contribution to the tensor-to-scalar ratio in single-field slow-roll inflation. The amplitude of the curvature power spectrum is $A_{\rm{s}}\approx2.2\times10^{-9}$ \cite{Ade:2015xua}.

In Figs. \ref{fig:EpsBrvace-3xi1M1}, \ref{fig:EpsBrvace-3xi10M10}, \ref{fig:EpsBrvace-3xi1M10}, \ref{fig:EpsBrvace-3xi10M1}, the orange solid line and the shaded area underneath it depict the inequality in Eq. \eqref{eq:SchwEpsB}, i.e., the regions where strong backreaction occurs due to the induced current. From this figures we can conclude that no additional constraint comes from the fermionic particle production.

Note that the contribution from chiral anomaly is small for the masses under consideration $\mu_m \gtrsim 1$, and also it is suppressed by a factor of $g_A^2$, (see \Cpageref{footnote2} for motivation behind this mass constraint, and see \cite{Domcke:2018gfr} for inclusion of one-loop effects for massless fermions).

\section{Discussion}\label{sec:Discussion}
\label{sec:concl}

We have studied the evolution of a Dirac field doublet, which is covariantly coupled to an axion and an isotropic $SU(2)$ gauge field background in de Sitter spacetime. We assumed the fermion field to have a Dirac mass term. Our work extends the previous work on fermion production from axion and abelian $U(1)$ gauge fields \cite{Hayashinaka:2016qqn, Adshead:2015kza, Adshead:2018oaa, Domcke:2018eki}, as well as on the simplest $SU(2)$ case with massless fermions \cite{Domcke:2018gfr}. 

We discovered that the $SU(2)$ background, in combination with the Dirac mass term, leads to non-trivial couplings between fermion components of different flavor and chirality. We then found a new convenient basis for the doublet of fermionic fields, given as a linear transformation in Fourier space of the original doublet, for which the action separates into two decoupled sectors. One of the sub-systems is solvable analytically, whereas the other sub sector is not and we solved it numerically.

Using these solutions, we computed the expectation values of the induced currents, which we used to estimate for what model parameters backreaction effects become important. More specifically, we considered the isotropic part of the $SU(2)$ matter current, as well as the 4-divergence of the axial current, which can be used to estimate the fermionic backreaction on the gauge field and axion backgrounds, respectively. 

To find the vacuum expectation values of bilinearies in fermionic fields, such as the currents, we had to deal with UV-divergent integrals. To this end, we extended the idea of an existing instantaneous vacuum subtraction scheme \cite{Figueroa:2013vif}, which involves the subtraction of the contribution of zero-point fluctuations to the currents. We extended it to fermionic models with most generic Hamiltonians, which permit only a numerical treatment. We compared the results with independent regularization scheme, i.e the point-splitting method and found excellent agreement for $S_{+}$ (the details of the latter method will appear in \cite{Maleknejad:2019hdr}). We find that the adiabatic vacuum subtraction scheme could not be utilized here, since there are instants where adiabatic modes are ill-behaved. We also made a careful distinction between the contributions of particles and antiparticles to the vacuum expectation values, which played an important role in the computation of the tree-level backreaction on the axion.

We showed that the $SU(2)$-background experiences strong backreaction due to fermions only for model parameters which are already excluded on observational and/or theoretical grounds (see Figs. $2-5$), similarly to the case of scalars \cite{Lozanov:2018kpk}. The tree-level expectation value of the 4-divergence of the axial current vanishes. We then estimated when the chiral anomaly, which is a loop effect, becomes important for the backreaction on the axion background. We find that backreaction remains unimportant provided that $f \gg H$.

We conclude that the background dynamics of an axion-$SU(2)$ gauge field spectator sector remains unaffected by production of fermions.

\acknowledgments
We thank Peter Adshead, Giovanni Cabass, Marco Peloso, Lorenzo Sorbo, and Yuki Watanabe for useful discussions. We are especially grateful to Eiichiro Komatsu for insightful discussions and comments on the manuscript.

\appendix

\section{Mathematical tools}\label{Math}

In this appendix, we review some mathematical tools that we use throughout this work. These include the definition of the direct sum, Kronecker product, the spin connection, and some asymptotic forms of Whittaker functions. 

The vector space $\bf{V}$, is the direct sum of two subspaces, $\bf{U}_1$ and $\bf{U}_2$, as \bea
\bf{V}= \bf{U}_1\oplus \bf{U}_2,
\eea
 if and only if
$\bf{V}=\bf{U}_1+\bf{U}_2$, and $\bf{U}_1$ and $\bf{U}_2$ are independent.

The Kronecker product of two matrices, ${\bf{A}}_{m\times n}$ and ${\bf{B}}_{q\times p}$, is defined as a $mp \times nq$ block matrix given by
\Beq\label{Kronecker}
{\bf{A}}\otimes {\bf{B}} = \begin{pmatrix}
A_{11} {\bf{B}} & \dots &  A_{1n}  {\bf{B}} \\
& \ddots & \\
A_{m1} {\bf{B}} & \dots & A_{mn}  {\bf{B}}
\end{pmatrix}.
\Eeq

The 8-spinor covariant derivative in \eqref{slashD} is
\Beq
D_{\mu}\otimes\gamma^{\alpha}\tilde{\Psi}\equiv\left(\I \nabla_{\mu}-ig_{\!A}{\bf{A}}_{\mu}\right) \otimes\gamma^{\alpha}\tilde{\Psi}\,,
\Eeq
where the spin covariant derivative is
\Beq\label{Spin-dee}
\nabla_{\mu}\tilde{\Psi} = [\II \partial_{\mu}+\omega_{\mu}] \tilde{\Psi},
\Eeq
with $\omega_{\mu}$ being the spin-connections
\bea
\omega_{\mu}=-\frac{i}{2} \omega_{\mu}^{~\alpha\beta} \sigma_{\alpha\beta},
\eea
and $\sigma_{\alpha\beta} = \frac{i}{4}[\gamma_{\alpha},\gamma_{\beta}]$ being the spinor generators of the Lorentz algebra. 
The elements of the spin-connection $\omega_{\mu}^{~\alpha\beta}$ are given by
\bea
\omega_{\mu}^{~\alpha\beta} = {\bf{e}}^{~\alpha}_{\nu} \nabla_{\mu} {\bf{e}}^{\nu\beta}.
\eea
In FLRW spacetime using the conformal time, the verbeins are
\Beq
{\bf{e}}^{\mu}_{~\alpha}=a(\tau)^{-1}\delta^{\mu}_{\alpha},
\Eeq
and the only non-zero components of the spin connection coefficients are
\Beq\label{Spin-connect-FRW}
 \omega_{\mu}^{~i0}=-\omega_{\mu}^{~0i}= -\mathcal{H}\delta^i_{\mu}.
\Eeq

Whittaker functions $W_{\kappa,\mu}(z)$ and $M_{\kappa,\mu}(z)$ take the following aymptotic forms in the limit $\mid z\mid\rightarrow\infty$

\Beq\label{WM-asymp}
W_{\kappa,\mu}(z) &\rightarrow z^{\kappa}e^{-z/2}, \\
M_{\kappa,\mu}(z) &\rightarrow \Gamma(2\mu+1)\bigg(\frac{i(-1)^{\mu-\kappa}z^{\kappa}e^{-z/2}}{\Gamma({-\kappa+\mu+\frac12})}\\&\qquad\qquad\qquad+\frac{z^{-\kappa}e^{z/2}}{\Gamma({-\kappa+\mu+\frac12})}\bigg),
\Eeq
where $\mid \arg z\mid<\frac32\pi$.
Thus, for a complex $\kappa$, we have
\Beq\label{general-BD}
\lim_{\tau\rightarrow -\infty} \frac{(2\x)^{-\kappa_{_{\rm{R}}}}}{\sqrt{2k}}e^{-\kappa_{_{\rm{I}}}\pi/2}W_{\kappa,\mu}(-2i\x) = \frac{1}{\sqrt{2k}} e^{-ik\tau},
\Eeq
where $\kappa_{_{\rm{R}}}$ and $\kappa_{_{\rm{I}}}$ are the real and imaginary parts of $\kappa$. Therefore, 
\bea
\frac{(2\x)^{-\kappa_{_{\rm{R}}}}}{(2\pi)^{3/2}\sqrt{2k}}e^{-\kappa_{_{\rm{I}}}\pi/2}W_{\kappa,\mu}(-2i\x)
\eea
represents the positive frequency solutions in the asymptotic past, i.e. the Bunch-Davies vacuum.

\section{The spinor subspaces, $\tilde\Uppsi=\Uppsi^+\oplus\Uppsi^-$}\label{helicity}

In this appendix, our aim is to reduce a 8-spinor into two 4-spinor. Again we recall that the tilde denotes $8\times 8$ matrices, $4\times 4$ matrices remained unaltered; and the $2\times 2$ matrices are presented in boldface. To find the spinor subspaces, it is convenient to apply the following chain of unitary transformations: First, in section \ref{subsec:b1} we review the Weyl representation, second, in section \ref{subsec:b2} we define the new extended helicity basis to transform $\Uppsi_L$ and $\Uppsi_R$ to $\Uppsi^+$ and $\Uppsi^-$, and in last section \ref{subsec:b3} we transform each of the subspinors to the Dirac representation.

\Beq
\tilde{\Uppsi}: \quad \underbrace{\begin{pmatrix}
\Uppsi^1 \\
\Uppsi^2
\end{pmatrix}}_{\text{flavor}} \quad  \xrightarrow{\tilde T_1}  \quad \underbrace{\begin{pmatrix}
\Uppsi_L \\
\Uppsi_R
\end{pmatrix}}_{\text{chiral}}  \quad \xrightarrow{\tilde T_2} \quad
 \underbrace{\begin{pmatrix}
\Uppsi^+ \\
\Uppsi^-
\end{pmatrix}}_{\text{extended helicity}}. \qquad \qquad
\Eeq

Note that the matrix $\tilde P_{\pm}(\bk)$ in the equation \eqref{eq:P} consists of all the above chain of unitary transformations.

\subsection{Weyl representation}
\label{subsec:b1}
The 8-spinor can be decomposed into two chirality states by the projection operators 
\Beq
\label{eq:proj}
\tilde P_{L,R}= \I \otimes \bigg(\frac{\II \mp\gamma^5}{2}\bigg),\,
\Eeq
such that left- and right-handed components are given by
\Beq
\tilde\Uppsi_{L,R}= \tilde P_{L,R}\tilde\Uppsi.
\Eeq
In \eqref{eq:proj} the minus and plus signs are for L and R, respectively. \\

The spinor in the flavor (Dirac) frame transforms into the chiral frame as
$$\tilde T_1 \begin{pmatrix}
\Uppsi^1 \\
\Uppsi^2
\end{pmatrix} = \begin{pmatrix}
\Uppsi_L \\
\Uppsi_R
\end{pmatrix},$$
where $\tilde T_1$ is the following unitary matrix
\bea
\tilde T_1 = \frac{1}{\sqrt{2}}\begin{pmatrix}
\I & -\I & \bf{0} & \bf{0} \\
\bf{0} & \bf{0} & \I & -\I \\
\I & \I & \bf{0} & \bf{0} \\
\bf{0} & \bf{0} & \I & \I
\end{pmatrix}.
\eea
Moreover, the $\I\otimes \gamma^{\alpha}$ operators transform as
\bea
\tilde{T}_1.(\I\otimes \gamma^{\alpha}). \tilde{T}_1^{-1} = \begin{pmatrix}
\bf{0} & \bf{0} & \boldsymbol{\sigma}^{\alpha} & \bf{0} \\
\bf{0} & \bf{0} & \bf{0} & \boldsymbol{\sigma}^{\alpha} \\
 \bar{\boldsymbol{\sigma}}^{\alpha} & \bf{0} & \bf{0} & \bf{0} \\
 \bf{0} & \bar{\boldsymbol{\sigma}}^{\alpha} & \bf{0} & \bf{0} \\
\end{pmatrix},
\eea
where 
\bea\label{sigma-mu}
\boldsymbol{\sigma}^\alpha = (\I, \boldsymbol{\sigma}^i) \an \bar{\boldsymbol{\sigma}}^\alpha = (\I, -\boldsymbol{\sigma}^i).
\eea
The (flat space) gamma matrices in the Weyl representation are
\bea
\gamma_{_{\rm{W}}}^{\alpha} = \begin{pmatrix}
\bold{0} & \boldsymbol{\sigma}^\alpha \\
\bar{\boldsymbol{\sigma}}^\alpha & \bold{0}
\end{pmatrix} \an  \gamma_{_{\rm{W}}}^5 = \begin{pmatrix}
-\I & \bold{0}  \\
 \bold{0} & \I
\end{pmatrix}.
\eea

In this frame, the action in \eqref{theory} is given as
\bea\label{chiral-A}
S = \int \rm{d}\tau {\rm{d}}k^3 (\bar{\Uppsi}_{R,\bf{k}} \bar{\Uppsi}_{L,\bf{k}}) ~ . ~ \tilde{\rm{L}}_{\bf{k}}(\tau )~.  \begin{pmatrix}
\Uppsi_{L,\bf{k}} \\
\Uppsi_{R,\bf{k}} 
\end{pmatrix},
\eea
where $\tilde{\rm{L}}_{\bf{k}}(\tau )$ is the following $8\times 8$ operator
\Beq\label{L-action}
\tilde{\rm{L}}_{\bf{k}}(\tau ) \equiv i \begin{pmatrix}
i\mu_{{\rm m}} \mH \II & \II \p_{\tau} + i \Sigma_4(\tau,\bk) \\
 \II \p_{\tau} - i\Sigma_4(\tau,\bk)  & i\mu_{{\rm m}} \mH  \II
\end{pmatrix},
\Eeq
where $\Sigma_4$ is the following $4\times 4$ operator
\Beq\label{Sigma4-}
\Sigma_{_{4}}(\tau,\bk) =  \I\otimes k^i.\boldsymbol{\sigma}^i + \mH (2\xpi {\II} - \frac{\xp}{2} \boldsymbol{\sigma}^i \otimes \boldsymbol{\sigma}^j \delta_{ij}).
\Eeq
In the absence of the mass term, the system in \eqref{chiral-A} is decomposed into two independent sub-sectors in terms of the left- and right-handed fields. However, in the massive case, we need to take one step further and expand the fields in the extended helicity states.

\subsection{Extended helicity states}
\label{subsec:b2}

The aim here is to write the $4\times 4$ operator, $\Sigma_4 $, in \eqref{L-action} as a block diagonal matrix. For that we need to find the common eigenstates of the $4\times 4$ helicity operator, $  \I\otimes k^i.\boldsymbol{\sigma}^i $, and $ \boldsymbol{\sigma}^i \otimes \boldsymbol{\sigma}_i$. These two have only 2 common eigenstates and therefore, it is not possible to fully diagonalize $\Sigma_4 $. Nevertheless, it can be block-diagonalized and decomposed into two subspaces.

For a given momentum, $k^{\alpha}$, the orthonormal eigenstates for the helicity operator, $ \I\otimes k^i.\boldsymbol{\sigma}^i $, are
\Beq
e^{+}_{~+}(k^{\alpha})&=\frac{\check{k}^{\alpha}\bar{\boldsymbol{\sigma}}_{\alpha}\otimes\check{k}^{\beta}\bar{\boldsymbol{\sigma}}_{\beta}}{2k(k+k^3)}\begin{pmatrix}1 \\ 0\\ 0 \\ 0\end{pmatrix}\,,\\
e^{+}_{~-}(k^{\alpha})&=-\frac{\check{k}^{\alpha}\boldsymbol{\sigma}_{\alpha}\otimes\check{k}^{\beta}\boldsymbol{\sigma}_{\beta}}{2k(k+k^3)}\begin{pmatrix}0 \\ 0\\ 0 \\ 1\end{pmatrix}\,,\\
e^{-}_{~+}(k^{\alpha})&=-\frac{\check{k}^{\alpha}\boldsymbol{\sigma}_{\alpha}\otimes\check{k}^{\beta}\bar{\boldsymbol{\sigma}}_{\beta}}{2k(k+k^3)}\begin{pmatrix}0 \\ 0\\ 1 \\ 0\end{pmatrix}\,, \\
e^{-}_{~-}(k^{\alpha})&=-\frac{\check{k}^{\alpha}\bar{\boldsymbol{\sigma}}_{\alpha}\otimes\check{k}^{\beta}\boldsymbol{\sigma}_{\beta}}{2k(k+k^3)}\begin{pmatrix}0 \\ 1\\ 0 \\ 0\end{pmatrix}\,,
\Eeq
where $\boldsymbol{\sigma}^{\alpha}$ and $\bar{\boldsymbol{\sigma}}^{\alpha}$ are given in \eqref{sigma-mu}, and their indices are lowered with the 
Minkowski metric, i.e. $\boldsymbol{\sigma}_{\alpha}= \eta_{\alpha\beta} \boldsymbol{\sigma}^{\beta}$, and 
 $\check{k}^{\alpha}$ is a four vector given as
\bea
\check{k}^{\alpha} \equiv (k,\bk),
\eea
where $k=\sqrt{k^i . k^i }$. Notice that $\check{k}^{\alpha}$ is the four momentum of the massless field, but for the massive cases it is just a mathematical tool.

The $ e^{p}_{~s}(k^{\alpha})$ with $p=\pm1$ and $s=\pm1$ satisfies the eigenstate equation 
\Beq
 \I\otimes k^i.\boldsymbol{\sigma}^i  e^{p}_{~s}(k^{\alpha}) = s ~ k e^{p}_{~s}(k^{\alpha})\,,
\Eeq
and the orthonormality condition
\bea
 \quad e^{p\dagger}_{~s}(k^{\alpha})\cdot e^{p'}_{~s'}(k^{\alpha})=\delta_{ss'}\delta^{pp'}\,.
\eea
The $p=+1$ objects are also eigenvectors of $\Sigma_{_{4}}(\tau,\bk)$ in \eqref{Sigma4-}
\Beq
\Sigma_{_{4}}(\tau,\bk)e^{+}_{~s}(k^{\alpha})=\lambda^+_{s}e^{+}_{~s}(k^{\alpha})\,,
\Eeq
where $\lambda^+_s=sk+(-\frac12\xi_{A}+ 2\xpi) \mH$.

Since $e^p_{~s}(k^{\alpha})$ make an orthonormal basis, we can expand fields and matrices in that frame by using the unitary transformation
\Beq\label{R-def}
R_{\bf{k}} &= \big[ e^{+}_{~+}(k^{\alpha}) ~  e^{+}_{~-}(k^{\alpha}) ~  e^{-}_{~-}(k^{\alpha}) ~  e^{-}_{~+}(k^{\alpha}) \big] \\
&=\begin{pmatrix*}[c]e^{+}_{~+1} & e^{+}_{~-1} & e^{-}_{~-1} & e^{-}_{~+1}\\ e^{+}_{~+2} & e^{+}_{~-2} & e^{-}_{~-2} & e^{-}_{~+2}\\ e^{+}_{~+3} & e^{+}_{~-3} & e^{-}_{~-3} & e^{-}_{~+3}\\e^{+}_{~+4} & e^{+}_{~-4} & e^{-}_{~-4} & e^{-}_{~+4}\end{pmatrix*},
\Eeq
where $e^p_{~si}$ is the $i$th element of the $e^p_{~s}(k^{\alpha})$.

For each given momentum, $\bk$, $R_{\bf{k}}$ takes the Weyl spinors to their helicity frame. The normal spinor basis in this frame are
\Beq
e_{L+} = \begin{pmatrix}
~{\bf{E}}_{+} \\
{\bf{0}}
\end{pmatrix}, \quad e_{L-} = \begin{pmatrix}
~{\bf{E}}_{-} \\
{\bf{0}}
\end{pmatrix}, \\
e_{R+} = \begin{pmatrix}
{\bf{0}} \\ 
~{\bf{E}}_{+}
\end{pmatrix}, \quad e_{R-} = \begin{pmatrix}
{\bf{0}} \\ 
~{\bf{E}}_{-}
\end{pmatrix},
\Eeq
where ${\bf{E}}_{+}$ and ${\bf{E}}_{-}$ are the following 2-spinor basis
\bea\label{E+-E-ap}
{\bf{E}}_{+} = \begin{pmatrix}
1 \\
0
\end{pmatrix} \an {\bf{E}}_{-} = \begin{pmatrix}
0 \\
1
\end{pmatrix}.
\eea

In the frame of helicity, $\Sigma_{_{4}}(\tau,\bk)$ takes a block diagonal form
\Beq
\label{eq:SigmaDiag}
\check{\Sigma}_{_{4}}(\tau,k)=R_{\bf{k}}^{-1}\Sigma_{_{4}}(\tau,\bk)R_{\bf{k}}=\begin{pmatrix*}[c]\check{\boldsymbol{\Sigma}}^+ & 0\\ 0 & \check{\boldsymbol{\Sigma}}^-\end{pmatrix*}\,,
\Eeq
where $\check{\boldsymbol{\Sigma}}^{\pm}$ are $2\times 2$ matrices defined as
\Beq
\label{eq:Sigmas}
&\check{\boldsymbol{\Sigma}}^+ \equiv k\boldsymbol{\sigma}^3+ (2\xpi-\frac{\xi_A}{2})\mathcal{H}\I \,,\\
&\check{\boldsymbol{\Sigma}}^- \equiv -k\boldsymbol{\sigma}^3 + (2\xpi + \frac{\xi_A}{2})\mathcal{H}\I  -\xi_A\mathcal{H}\boldsymbol{\sigma}^1\,.
\Eeq

The 8-spinor in the Weyl representation can be written in terms of the helicity states as
\bea\label{Rspinor}
\tilde R_{\bf{k}} \begin{pmatrix}
\Uppsi_{L,\bf{k}} \\
\Uppsi_{R,\bf{k}} 
\end{pmatrix} = \begin{pmatrix}
\boldsymbol{\Uppsi}^+_{L,\bf{k}} \\
\boldsymbol{\Uppsi}^-_{L,\bf{k}} \\
\boldsymbol{\Uppsi}^+_{R,\bf{k}} \\
\boldsymbol{\Uppsi}^-_{R,\bf{k}}
\end{pmatrix},
\eea
where $\tilde R_{\bf{k}}$ is the following $8\times 8$ unitary operator 
\bea
\tilde R_{\bf{k}} \equiv \I \otimes R_{\bf{k}}^{-1}.
\eea
The Lagrangian operator in \eqref{L-action} is also transformed as
\Beq
 \check L_{\bf{k}}(\tau) &= \tilde R_{\bf{k}} ~ . ~ \tilde L_{\bf{k}}(\tau) ~.~ \tilde R_{\bf{k}}^{-1} \\
&= i \begin{pmatrix}
i\mu_{{\rm m}} \mH \II & \II \p_{\tau} + i \check\Sigma_4(\tau,\bk) \\
 \II \p_{\tau} - i\check\Sigma_4(\tau,\bk)  & i\mu_{{\rm m}} \mH  \II
\end{pmatrix},
\Eeq
which decouples the plus and minus spinors in \eqref{Rspinor}. This split would be clearer if we take another unitary transformation 
\bea
\tilde S =  \begin{pmatrix}
\I & 0 & 0 & 0\\
0 & 0 & \I & 0 \\
0 & \I & 0 & 0 \\
0 & 0 & 0 & \I
\end{pmatrix},
\eea 
and define $\tilde T_{2,\bf{k}}$ as
\bea
\tilde T_{2,\bf{k}} \equiv \tilde S \tilde R_{\bf{k}}.
\eea
Under the action of $\tilde T_{2,\bf{k}}$, the 8-spinor transforms as
\bea\label{pmspinor}
\tilde T_{2,\bf{k}} \begin{pmatrix}
\Uppsi_{L,\bf{k}} \\
\Uppsi_{R,\bf{k}} 
\end{pmatrix} = \begin{pmatrix}
\Uppsi^+_{\bf{k}} \\
\Uppsi^-_{\bf{k}}
\end{pmatrix},
\eea
and the Lagrangian operator becomes the following block diagonal $8\times 8$ matrix
\Beq
 L_{\bf{k}}(\tau) = \tilde T_{2,\bf{k}} ~ . ~ \tilde L_{\bf{k}}(\tau) ~.~ \tilde T_{2,\bf{k}}^{-1} =  \begin{pmatrix}
L^+_{\bf{k}}(\tau) & 0 \\
0 & L^-_{\bf{k}}(\tau)
\end{pmatrix},
\Eeq
where $L^{\pm}_{\bf{k}}(\tau)$ are the following $4\times 4$ operations
\Beq
L^{\pm}_{\bf{k}}(\tau) \equiv i \begin{pmatrix}
i\mu_{{\rm m}} \mH \I & \I \p_{\tau} + i \boldsymbol{\check{\Sigma}}^{\pm}(\tau,\bk) \\
 \I \p_{\tau} - i\boldsymbol{\check{\Sigma}}^{\pm}(\tau,\bk)  & i\mu_{{\rm m}} \mH  \I
\end{pmatrix}.
\Eeq
Here, $\boldsymbol{\check{\Sigma}}^{\pm}(\tau,\bk)$ are given in \eqref{eq:Sigmas}. Therefore, the theory in \eqref{chiral-A} splits into two subsectors in terms of the plus and minus spinors as
\bea
S[\tilde\Uppsi] = S_+[\tilde\Uppsi^+] + S_-[\tilde\Uppsi^-],
\eea
where 
\bea\label{chiral-Ah}
S_{\pm} = \int \frac{{\rm{d}}\tau {\rm{d}}k^3}{(2\pi)^3} \bar{\Uppsi}^{\pm}_{\bf{k}} ~ . ~ \tilde{\rm{L}}^{\pm}_{\bf{k}}(\tau )~.  
\Uppsi^{\pm}_{\bf{k}}.
\eea
The $\tilde{\rm{L}}^{\pm}_{\bf{k}}(\tau )$ operators are given as
\Beq\label{eq:L+W}
\tilde{\rm{L}}^{+}_{\bf{k}}(\tau ) & \equiv  \left[i\gamma^0_{_{\rm{W}}} \partial_{\tau} - k\gamma^3_{_{\rm{W}}} -\left(2\xpi -\frac{\xi_A}{2}\right)\mathcal{H}\lambda_{_{4}} - \mu_{{\rm m}}\mH \II \right]\,,
\Eeq
\Beq\label{eq:L-W}
\tilde{\rm{L}}^{-}_{\bf{k}}(\tau ) \equiv  \Bigg[i\gamma^0_{_{\rm{W}}} \partial_{\tau}+k \gamma^3_{_{\rm{W}}} +\gamma^1_{_{\rm{W}}}\xi_A\mathcal{H}-\big(2\xpi &+\frac{\xi_A}{2}\big)\mathcal{H}\lambda_{_{4}} \\
&-\mu_{{\rm m}} \mH \II \Bigg],
\Eeq
where $\gamma^{\alpha}_{_{\rm{W}}}$s are the gamma matrices in the Weyl representation and $\lambda_{_{\rm{W}}}$ is
\bea
\lambda_{_{4}} \equiv \begin{pmatrix}
0 & \I \\
-\I & 0
\end{pmatrix}.
\eea
We thus showed that our 8-spinor space splits into two irreducible representations
\bea
\tilde \Uppsi = \Uppsi^+ \oplus \Uppsi^-,
\eea 
in terms of two 4-spinors $\Uppsi^+$ and $\Uppsi^-$.

\subsection{Sub-spinors in Dirac frame}
\label{subsec:b3}

Up to this point, we have split the spinor space into two subspaces but each of the supspinors are still in their Weyl representation (see \eqref{pmspinor})
\bea\label{pmspinor-D}
\tilde V_{\bf{k}} \begin{pmatrix}
\Uppsi_{L,\bf{k}} \\
\Uppsi_{R,\bf{k}} 
\end{pmatrix} = \begin{pmatrix}
{\boldsymbol{\Uppsi}}^+_{L,\bf{k}} \\
{\boldsymbol{\Uppsi}}^+_{R,\bf{k}} \\
{\boldsymbol{\Uppsi}}^-_{L,\bf{k}} \\
{\boldsymbol{\Uppsi}}^-_{R,\bf{k}}
\end{pmatrix}.
\eea
The 4-spinors can be transformed to their Dirac representation as
\bea
\Uppsi^{\pm}_{\rm{D}} = \mathcal{D} ~ \Uppsi^{\pm},
\eea
where
\bea
\mathcal{D} = \frac{1}{\sqrt{2}} \begin{pmatrix}
\I & \I \\
-\I & ~~\I
\end{pmatrix}.
\eea
The gamma matrices in the Weyl representation, $\gamma^{\alpha}_{_{\rm{W}}}$, and Dirac representation, $\gamma^{\alpha}$, are related as $\gamma^{\alpha} = \mathcal{D} \gamma^{\alpha}_{_{\rm{W}}} \mathcal{D}^{-1}$.
Moreover, the Lagrangian operators of each of the subspaces in \eqref{eq:L+W} and \eqref{eq:L-W} take the following forms in the Dirac representation
\Beq
\tilde{\rm{L}}^{+}_{\bf{k},\rm{D}}(\tau ) & \equiv \left[i\gamma^0 \partial_{\tau} - k\gamma^3 -\left(2\xpi -\frac{\xi_A}{2}\right)\mathcal{H}\lambda_{_{4}} - \mu_{{\rm m}}\mH \II \right]\,,\\
\tilde{\rm{L}}^{-}_{\bf{k},\rm{D}}(\tau ) & \equiv \Bigg[i\gamma^0 \partial_{\tau}+k \gamma^3 +\gamma^1\xi_A\mathcal{H}-\left(2\xpi +\frac{\xi_A}{2}\right)\mathcal{H}\lambda_{_{4}} \\
&\qquad\qquad\qquad\qquad\qquad\qquad-\mu_{{\rm m}} \mH \II \Bigg].
\Eeq

In the Dirac-helicity frame which we introduced above, the Dirac fields can be expanded as
\bea
\Uppsi^{\pm}_{\bf{k}} = \sum_{s=\pm} \begin{pmatrix}
~\uppsi^{\pm\uparrow}_{s}(\tau,k) {\bf{E}}_s \\
\\
s \uppsi^{\pm\downarrow}_{s}(\tau,k)  {\bf{E}}_s
\end{pmatrix},
\eea
where $\uppsi^{\pm\uparrow}_{s}(\tau,k)$ and $\uppsi^{\pm\downarrow}_{s}(\tau,k)$ are mode functions and ${\bf{E}}_s$ with $s=\pm 1$ are the two-spinor polarization states given in \eqref{E+-E-ap}.

Notice that $\gamma^{\alpha}$ are the gamma functions in the Dirac representation and $\lambda_4$ is the same in both Weyl and Dirac representations. For the sake of simplicity, in the main text we remove the $D$ subscript and write the fields and operators in the Dirac representation.

\section{Hamiltonian}\label{Hamiltonian}

In this appendix we derive and diagonalize the Hamiltonian of our model. 
The Hamiltonian is derived from the actions, given in Eq. \eqref{eq:SLSR-P} and \eqref{eq:SLSR-M}, by defining the Lagrangian $S_{\pm}\equiv\int {\rm d}\tau L_{\pm}$ and then carrying out a Legendre transformation.  We derive the Hamiltonians for $S_+$ and $S_-$ separately and diagonalize them afterwards. \\
Before we calculate the Hamiltonian, we need to explain the quantization procedure for the fermions in $S_+$ and $S_-$. 
\subsection{Quantization of the $S_+$ fermions}

The quantization procedure for the $\Uppsi ^{+}_{\bk}$ modes is the following. We first define the canonical conjugate momenta
\Beq
\pi^{\Uppsi +}_{\bk,\alpha}=\frac{\delta S_+}{\delta \partial_{\tau}\Uppsi ^{+}_{\bk,\alpha}}=i\Uppsi _{+,\bk,\alpha}^{*}\,,
\Eeq
where $\alpha$ runs from $1$ to $4$.
We then promote $\Uppsi ^{+}_{\bk}$ and $\pi^{\Uppsi +}_{\bk}$ to quantum operators, obeying the canonical equal-time anti-commutation relations
\Beq
\label{eq:AnticommFields}
&\{\Uppsi ^{+}_{\bk,\alpha}(\tau),\Uppsi ^{+}_{\bk',\beta}(\tau)\}=0\,,\\
&\{\pi^{\Uppsi+}_{\bk,\alpha}(\tau),\pi^{\Uppsi+}_{\bk',\beta}(\tau)\}=0\,,\\
&\{\Uppsi^+_{\bk,\alpha}(\tau),\pi^{\Uppsi+}_{\bk',\beta}(\tau)\}=i(2\pi)^{-3}\delta_{\alpha\beta}\delta^{(3)}(\bk-\bk')\,.
\Eeq

We also impose that the time-independent coefficients in Eq. \eqref{eq:psipls} are the standard anti-commuting creation and annihilation operators, i.e.,
\Beq
\label{eq:Anticommab}
\{a^{+}_{s,\bk},a^{+\dagger}_{s',\bk'}\}&=\delta_{ss'}\delta^{(3)}(\bk-\bk')\,, \\
\{b^{+}_{s,\bk},b^{+\dagger}_{s',\bk'}\}&=\delta_{ss'}\delta^{(3)}(\bk-\bk')\,,
\Eeq
with all other anti-commutators vanishing. The canonical quantization expressions in Eqs. (\ref{eq:AnticommFields},\ref{eq:Anticommab}) yield the following normalization condition
\Beq
\label{eq:NormCond}
&\sum_{s=\pm}\Bigg[({U}^{+}_{s,k}(\tau))_{\alpha}({U}^{+\dagger}_{s,k}(\tau))_{\beta}\\
&\qquad+({V}^{+}_{s,k}(\tau))_{\alpha}({V}^{+\dagger}_{s,k}(\tau))_{\beta}\Bigg]=\delta_{\alpha\beta}(2\pi)^{-3}\,.
\Eeq

One can check that each term in the square brackets is indeed a constant, i.e., preserved by the equation of motion given in Eq. \eqref{EOM-psi-p}. In the following section each constant is determined after assuming that at very early times the modes are in the Bunch-Davies vacuum, i.e., 
\Beq
\label{eq:BDinitconds}
\lim\limits_{k\tau\to-\infty}{U}_{s,k}^{+}(\tau)\propto e^{-ik\tau}\,, \qquad \lim\limits_{k\tau\to-\infty}{V}_{s,k}^{+}(\tau)\propto e^{ik\tau}\,,
\Eeq
corresponding to the positive and negative frequency solutions, respectively. In addition to that, the amplitudes and the relative phases of the Bunch-Davies vacuum modes are such that the corresponding Hamiltonian is diagonalized.
\subsection{$S_+$ Hamiltonian}

For $S_+$, 
\Beq
H_+&=\int {\rm d}^3k \left(\pi^{\Uppsi+}_{\bk,\alpha}\partial_{\tau}\Uppsi^+_{\bk,\alpha}\right)-L_+\\
&=\int\text{d}^3k{\Uppsi}^{+,\dagger}_{\bk}\gamma^0\Bigg[\gamma^3k+\left(2\xi_{\varphi} -\frac{\xi_A}{2}\right)\mathcal{H}\lambda_4 \\
&\qquad\qquad\qquad\qquad\qquad\qquad+\mu_{m}\mathcal{H}{\rm I}_{4}\Bigg] \Uppsi^+_{\bk}\,.
\Eeq
Using the mode function expansion from Eqs. (\ref{eq:psipls}, \ref{eq:UVpsipls}) in the Hamiltonian, we arrive at
\Beq
H_+&=\int{\rm d}^3k\sum_{s=\pm}\frac{1}{2}(a_{s,\bk}^{+\dagger},b_{s,-\bk}^{+})\\
&\begin{pmatrix}E\big(u^{\uparrow}_s,u^{\downarrow}_s\big) & F^*\big(u^{\uparrow,\downarrow}_s,v^{\uparrow,\downarrow}_s\big) \\ F\big(u^{\uparrow,\downarrow}_s,v^{\uparrow,\downarrow}_s\big) & E\big(v^{\uparrow}_s,v^{\downarrow}_s\big)\end{pmatrix}\begin{pmatrix}a_{s,\bk}^{+}\\b_{s,-\bk}^{+\dagger}\end{pmatrix}\,,
\Eeq
where
\Beq
\label{eq:EF}
&E\big(u^{\uparrow}_s,u^{\downarrow}_s\big)=2\left[k+s\left(2\xi_{\varphi}-\frac{\xi_A}{2}\right)\mathcal{H}\right]\Re\big(u^{\uparrow*}_{s}u^{\downarrow}_{s}\big)\\
&\qquad\qquad\qquad\qquad\qquad+\mu_{m}\mathcal{H}\left(|u^{\uparrow}_{s}|^2-|u^{\downarrow}_{s}|^2\right)\,,\\
&F\big(u^{\uparrow,\downarrow}_s,v^{\uparrow,\downarrow}_s\big)=\left[k+s\left(2\xi_{\varphi}-\frac{\xi_A}{2}\right)\mathcal{H}\right]\big(u_{s}^{\downarrow}v_{s}^{\uparrow*}+u_{s}^{\uparrow}v_{s}^{\downarrow*}\big)\\
&\qquad\qquad\qquad\qquad\qquad+\mu_{m}\mathcal{H}\left(v^{\uparrow*}_{s}u^{\uparrow}_{s}-v^{\downarrow*}_{s}u^{\downarrow}_{s}\right)\,.
\Eeq

To bring the Hamiltonian into a diagonal form we make a time-dependent Bogoliubov transformation
\Beq
\begin{pmatrix}\check{a}_{s,\bk}^{+}(\tau)\\ \check{b}_{s,-\bk}^{+\dagger}(\tau)\end{pmatrix}=\begin{pmatrix}\alpha_{s,k}(\tau) & \beta_{s,k}(\tau) \\ -\beta^*_{s,k}(\tau) & \alpha^*_{s,k}(\tau) \end{pmatrix}\begin{pmatrix}a_{s,\bk}^{+}\\b_{s,-\bk}^{+\dagger}\end{pmatrix}\,.
\Eeq
The new set of time-dependent creation and annihilation operators, $\check{a}_{s,\bk}^{+}(\tau)$ and $\check{b}_{s,-\bk}^{+\dagger}(\tau)$, respect the canonical anti-commutation relations, given in Eq. \eqref{eq:Anticommab}, iff the Bogoliubov coefficients satisfy
\Beq
|\alpha_{s,k}(\tau)|^2+|\beta_{s,k}(\tau)|^2=1\,.
\Eeq
This condition is met and the Hamiltonian is diagonalized as
\Beq
H_+=\int{\rm d}^3k\sum_{s=\pm}\left[\check{a}_{s,\bk}^{+\dagger}(\tau)\check{a}_{s,\bk}^{+}(\tau)-\check{b}_{s,-\bk}^{+}\check{b}_{s,-\bk}^{+\dagger}(\tau)\right]\omega_{s,k}(\tau)\,,
\Eeq
for
\Beq
\label{eq:alphabeta}
&|\beta_{s,k}(\tau)|^2=\frac{1}{2}\\
\times&\left[1-\frac{E\big(u^{\uparrow}_s,u^{\downarrow}_s\big)-E\big(v^{\uparrow}_s,v^{\downarrow}_s\big)}{\sqrt{4|F\big(u^{\uparrow,\downarrow}_s,v^{\uparrow,\downarrow}_s\big)|^2+(E\big(u^{\uparrow}_s,u^{\downarrow}_s\big)-E\big(v^{\uparrow}_s,v^{\downarrow}_s\big))^2}}\right]\,,\\
&|\alpha_{s,k}(\tau)|^2=\frac{1}{2}\\
\times&\left[1+\frac{E\big(u^{\uparrow}_s,u^{\downarrow}_s\big)-E\big(v^{\uparrow}_s,v^{\downarrow}_s\big)}{\sqrt{4|F\big(u^{\uparrow,\downarrow}_s,v^{\uparrow,\downarrow}_s\big)|^2+(E\big(u^{\uparrow}_s,u^{\downarrow}_s\big)-E\big(v^{\uparrow}_s,v^{\downarrow}_s\big))^2}}\right]\,,\\
&\alpha_{s,k}(\tau)=|\alpha_{s,k}(\tau)|e^{i\phi_F}\,,\quad \beta_{s,k}(\tau)=|\beta_{s,k}(\tau)|e^{-i\phi_F}\,,\\
&F\big(u^{\uparrow,\downarrow}_s,v^{\uparrow,\downarrow}_s\big)=|F\big(u^{\uparrow,\downarrow}_s,v^{\uparrow,\downarrow}_s\big)|e^{2i\phi_F}\,.
\Eeq
The effective frequency is given by
\Beq
\label{eq:omegagenpls}
&\omega_{s,k}(\tau)=\frac{E\big(u^{\uparrow}_s,u^{\downarrow}_s\big)+E\big(v^{\uparrow}_s,v^{\downarrow}_s\big)}{4}\\
&+\frac{1}{4}\sqrt{4|F\big(u^{\uparrow,\downarrow}_s,v^{\uparrow,\downarrow}_s\big)|^2+(E\big(u^{\uparrow}_s,u^{\downarrow}_s\big)-E\big(v^{\uparrow}_s,v^{\downarrow}_s\big))^2}\,.
\Eeq

It is important to note that the amplitudes and the relative phases of the Bunch-Davies vacuum modes are such that the corresponding Hamiltonian is diagonalized. The following analysis will fix our initial conditions for the positive and negative frequency solutions. \\ 

The Bunch-Davies vacuum is defined as
\Beq
a_{s,\bk}^{+}|0_{BD}\rangle=0\,,\quad b_{s,\bk}^{+}|0_{BD}\rangle=0\,,
\Eeq
whereas the instantaneous (or quasi-particle) vacuum as
\Beq
\check{a}_{s,\bk}^{+}(\tau)|0_{\tau}\rangle=0\,,\quad \check{b}_{s,\bk}^{+}(\tau)|0_{\tau}\rangle=0\,.
\Eeq
We work in the Heisenberg picture and we assume that the Universe is in the Bunch-Davies vacuum. The expectation values of observables are calculated with respect to it, i.e., the expected particle number is given by
\Beq
\check{N}_{s,\bk}(\tau)=&\langle0_{BD}|\check{n}_{s,\bk}(\tau)|0_{BD}\rangle\\
=&\langle0_{BD}|\check{a}^{+\dagger}_{s,\bk}(\tau)\check{a}^{+}_{s,\bk}(\tau)|0_{BD}\rangle=|\beta_{s,k}(\tau)|^2\,.
\Eeq
Hence, $|\beta_{s,k}(\tau)|^2$ is the occupation number of particles with given $s$ and $\bk$ at a time $\tau$.

We assume that at very early times, $k\tau\rightarrow-\infty$, $\Uppsi^+_{\bk}$ starts in the Bunch-Davies vacuum, i.e., its particle occupation numbers vanish
\Beq
\label{eq:betainit}
\lim\limits_{k\tau\to-\infty}{\beta}_{s,k}(\tau)=0\,.
\Eeq 
Then it follows from Eq. \eqref{eq:alphabeta} that $\lim\limits_{k\tau\to-\infty}F\big(u^{\uparrow,\downarrow}_s,v^{\uparrow,\downarrow}\big)=0$, which is satisfied, according to Eq. \eqref{eq:EF}, if $\lim\limits_{k\tau\to-\infty}(u^{\uparrow}_{s}(k,\tau)-v^{\uparrow*}_{s}(k,\tau))=0$ and $\lim\limits_{k\tau\to-\infty}(u^{\downarrow}_{s}(k,\tau)+v^{\downarrow*}_{s}(k,\tau))=0$. Note that there is some residual freedom in choosing the latter such that the Hamiltonian is diagonalized. One can show that the last two conditions are preserved by the equations of motion, i.e., if imposed initially they hold at later times (for arbitrary $\tau$) as well
\Beq
u^{\uparrow}_{s}(k,\tau)=v^{\uparrow*}_{s}(k,\tau)\quad {\rm and} \quad u^{\downarrow}_{s}(k,\tau)=-v^{\downarrow*}_{s}(k,\tau)\,.
\Eeq
Eqs. (\ref{eq:EF},\ref{eq:alphabeta},\ref{eq:omegagenpls}) then yield (for all $\tau$)
\Beq
&E\big(u^{\uparrow}_s,u^{\downarrow}_s\big)=-E\big(v^{\uparrow}_s,v^{\downarrow}_s\big)\,,\\
&|\beta_{s,k}(\tau)|^2=\frac{1}{2}\left[1-\frac{E\big(u^{\uparrow}_s,u^{\downarrow}_s\big)}{2\omega_{s,k}(\tau)}\right]\,,\\
&|\alpha_{s,k}(\tau)|^2=\frac{1}{2}\left[1+\frac{E\big(u^{\uparrow}_s,u^{\downarrow}_s\big)}{2\omega_{s,k}(\tau)}\right]\,,
\Eeq
where the effective frequency has been simplified to
\Beq
\omega_{s,k}(\tau)=\frac{1}{2}\sqrt{|F|^2+E\big(u^{\uparrow}_s,u^{\downarrow}_s\big)^2}\,.
\Eeq
The last condition one has to impose for Eq. \eqref{eq:betainit} to hold is $\lim\limits_{k\tau\to-\infty}\Re\big(u^{\uparrow*}_{s}(k,\tau)u^{\downarrow}_{s}(k,\tau)\big)=1$. \\
Note that then in the Bunch-Davies limit $\lim\limits_{k\tau\to-\infty}\omega_{s,k}=k$.

Therefore, after applying Eqs. (\ref{eq:NormCond},\ref{eq:BDinitconds}) and
\Beq
&\lim\limits_{k\tau\to-\infty}(u^{\uparrow}_{s}(k,\tau)-v^{\uparrow*}_{s}(k,\tau))=0\,,\\
&\lim\limits_{k\tau\to-\infty}(u^{\downarrow}_{s}(k,\tau)+v^{\downarrow*}_{s}(k,\tau))=0\,,\\
&\lim\limits_{k\tau\to-\infty}\Re\big(u^{\uparrow*}_{s}(k,\tau)u^{\downarrow}_{s}(k,\tau)\big)=1\,,
\Eeq
to Eqs. \eqref{eq:uupdowngen} and \eqref{eq.Ys-Zs} we can completely fix the solutions for the $u^{\uparrow,\downarrow}_{s}(k,\tau)$ mode functions, whereas the solutions for $v^{\uparrow,\downarrow}_{s}(k,\tau)$ follow from the Eq. \eqref{eq:uvpl}.
We also made use of the asymptotic form of the Whittaker function in (\ref{WM-asymp}).

\subsection{Quantization of the $S_-$ fermions}

The quantization prescription for the $\Uppsi ^{-}_{\bk}$ modes remains unchanged. It begins with the definition of the canonical conjugate momenta
\Beq
\pi^{\Uppsi -}_{\bk,\alpha}=\frac{\delta S_-}{\delta \partial_{\tau}\Uppsi ^{-}_{\bk,\alpha}}=i \Uppsi _{\bk,\alpha}^{-,*}\,,
\Eeq
where $\alpha$ runs from $1$ to $4$.
Then $\Uppsi ^{-}_{\bk}$ and $\pi^{\Uppsi -}_{\bk}$ are promoted to quantum operators, satisfying the canonical equal-time anti-commutation relations
\Beq
\label{eq:AnticommFieldsm}
&\{\Uppsi^-_{\bk,\alpha}(\tau),\Uppsi^-_{\bk',\beta}(\tau)\}=0\,,\\
&\{\pi^{\Uppsi-}_{\bk,\alpha}(\tau),\pi^{\Uppsi-}_{\bk',\beta}(\tau)\}=0\,,\\
&\{\Uppsi^-_{\bk,\alpha}(\tau),\pi^{\Uppsi-}_{\bk',\beta}(\tau)\}=i(2\pi)^{-3}\delta_{\alpha\beta}\delta^{(3)}(\bk-\bk')\,.
\Eeq
We again postulate that the time-independent coefficients in Eq. \eqref{eq:psimns} are the standard anti-commuting creation and annihilation operators, i.e.,
\Beq
\label{eq:Anticommabm}
\{a^{-}_{s,\bk},a^{-\dagger}_{s',\bk'}\}=\delta_{ss'}\delta^{(3)}(\bk-\bk')\,, \\
\{b^{-}_{s,\bk},b^{-\dagger}_{s',\bk'}\}=\delta_{ss'}\delta^{(3)}(\bk-\bk')\,,
\Eeq
with the rest of the anti-commutators vanishing. Eqs. (\ref{eq:AnticommFieldsm},\ref{eq:Anticommabm}) then imply the normalization condition
\Beq
\label{eq:NormCondm}
\sum_{s=\pm}&\Bigg[({U}^{-}_{s,k}(\tau))_{\alpha}({U}^{-\dagger}_{s,k}(\tau))_{\beta}\\
&+({V}^{-}_{s,k}(\tau))_{\alpha}({V}^{-\dagger}_{s,k}(\tau))_{\beta}\Bigg]=\delta_{\alpha\beta}(2\pi)^{-3}\,.
\Eeq
Every term in the square brackets is constant, according to the equation of motion given in Eq. \eqref{EOM-psi-m}. To find the constants, we again assume that the early-time modes are in the Bunch-Davies vacuum 
\Beq
\label{eq:BDinitcondsm}
\lim\limits_{k\tau\to-\infty}{U}_{s,k}^{-}(\tau)\propto e^{-ik\tau}\,, \qquad \lim\limits_{k\tau\to-\infty}{V}_{s,k}^{-}(\tau)\propto e^{ik\tau}\,.
\Eeq
Furthermore, the amplitudes and the phases of the modes are adjusted to diagonalize the Hamiltonian. 

\subsection{$S_-$ Hamiltonian}
For $S_-$, 
\Beq
H_-&=\int {\rm d}^3k \left(\pi^{\Uppsi-}_{\bk,\alpha}\partial_{\tau}\Uppsi^-_{\bk,\alpha}\right)-L_-\\
&=\int\text{d}^3k{\Uppsi}^{-,\dagger}_{\bk}\gamma^0\Bigg[-\gamma^3k-\gamma^1\xi_A\mathcal{H}\\
&\qquad+\left(2\xi_{\varphi} +\frac{\xi_A}{2}\right)\mathcal{H}\lambda_{4}+\mu_{m}\mathcal{H}{\rm I}_{4}\Bigg]{\Uppsi}^-_{\bk}\,.
\Eeq
After using the mode function expansion from Eqs. (\ref{eq:psimns},\ref{eq:UVpsimns}) in the Hamiltonian, we get
\Beq
H_-=\int\frac{{\rm d}^3k}{2}({\bf a}_{\bk}^{-\dagger},{\bf b}_{-\bk}^{-})\begin{pmatrix}{\bf E}^u & {\bf F}^{\dagger} \\ {\bf F} & {\bf E}^v\end{pmatrix}\begin{pmatrix}{\bf a}_{\bk}^{-}\\{\bf b}_{-\bk}^{-\dagger}\end{pmatrix}\,,
\Eeq
where
\Beq
{\bf a}_{\bk}^{-}=\begin{pmatrix}a_{+,\bk}^{-}\\a_{-,\bk}^{-}\end{pmatrix}\,,\quad {\bf b}_{-\bk}^{-\dagger}=\begin{pmatrix}b_{+,-\bk}^{-\dagger}\\b_{-,-\bk}^{-\dagger}\end{pmatrix}\,,\\
{\bf E}^u=\begin{pmatrix}E^u_+ & E^{u*}_{{\rm mix}}\\  E^{u}_{{\rm mix}} & E^u_-\end{pmatrix}\,,\quad {\bf F}=\begin{pmatrix}F_+ & F_{-,{\rm mix}}\\F_{+,{\rm mix}} & F_-\end{pmatrix}\,,\\
\Eeq
and
\Beq
\label{eq:EFm}
&E^u_s=\sum_{p=\pm}\bigg\{\!-2\left[k+sp\left(2\xi_{\varphi}+\frac{\xi_A}{2}\right)\mathcal{H}\right]\Re\big(u^{\uparrow*}_{s,p}u^{\downarrow}_{s,p}\big)\\
&+\mu_{m}\mathcal{H}\left(|u^{\uparrow}_{s,p}|^2-|u^{\downarrow}_{s,p}|^2\right)+2sp\xi_A\mathcal{H}\Re\big(u^{\uparrow*}_{s,p}u^{\downarrow}_{s,-p}\big)\bigg\}\,,\\
&F_s=\sum_{p=\pm}\bigg\{\!-\left[k+sp\left(2\xi_{\varphi}+\frac{\xi_A}{2}\right)\mathcal{H}\right]\big(v_{s,p}^{\uparrow*}u_{s,p}^{\downarrow}+v_{s,p}^{\downarrow*}u_{s,p}^{\uparrow}\big)\\
&+\mu_{m}\mathcal{H}\left(v^{\uparrow*}_{s,p}u^{\uparrow}_{s,p}-v^{\downarrow*}_{s,p}u^{\downarrow}_{s,p}\right)\\
&+sp\xi_A\mathcal{H}\left(v^{\uparrow*}_{s,p}u^{\downarrow}_{s,-p}+u^{\uparrow}_{s,p}v^{\downarrow*}_{s,-p}\right)\bigg\}\,,\\
&E^u_{{\rm mix}}=\sum_{p=\pm}\bigg\{\!-\left[k+p\left(2\xi_{\varphi}+\frac{\xi_A}{2}\right)\mathcal{H}\right]\big(u_{-,-p}^{\uparrow*}u_{+,p}^{\downarrow}\\
&+u_{-,-p}^{\downarrow*}u_{+,p}^{\uparrow}\big)+\mu_{m}\mathcal{H}\left(u^{\uparrow*}_{-,-p}u^{\uparrow}_{+,p}-u^{\downarrow*}_{-,-p}u^{\downarrow}_{+,p}\right)\\
&+p\xi_A\mathcal{H}\left(u^{\uparrow*}_{-,-p}u^{\downarrow}_{+,-p}+u^{\uparrow}_{+,p}u^{\downarrow*}_{-,p}\right)\bigg\}\,,\\
&F_{s,{\rm mix}}=\sum_{p=\pm}\bigg\{\!-\left[k+sp\left(2\xi_{\varphi}+\frac{\xi_A}{2}\right)\mathcal{H}\right]\big(v_{-s,-p}^{\uparrow*}u_{s,p}^{\downarrow}\\
&+v_{-s,-p}^{\downarrow*}u_{s,p}^{\uparrow}\big)+\mu_{m}\mathcal{H}\left(v^{\uparrow*}_{-s,p}u^{\uparrow}_{s,-p}-v^{\downarrow*}_{-s,p}u^{\downarrow}_{s,-p}\right)\\
&+sp\xi_A\mathcal{H}\left(v^{\uparrow*}_{-s,-p}u^{\downarrow}_{s,-p}+u^{\uparrow}_{s,p}v^{\downarrow*}_{-s,p}\right)\bigg\}\,.
\Eeq
The Hamiltonian can be diagonalized after making a time-dependent Bogoliubov transformation
\Beq
\begin{pmatrix}\check{\bf{a}}_{\bk}^{-}(\tau)\\ \check{\bf{b}}_{-\bk}^{-\dagger}(\tau)\end{pmatrix}=P_k(\tau)\begin{pmatrix}{\bf a}_{\bk}^{-}\\{\bf b}_{-\bk}^{-\dagger}\end{pmatrix}\,.
\Eeq
The time-dependent creation and annihilation operators, $\check{\bf a}_{\bk}^{-}(\tau)$ and $\check{\bf b}_{-\bk}^{-\dagger}(\tau)$, obey the canonical anti-commutation relations from Eq. \eqref{eq:Anticommabm}, iff the transformation matrix is unitary
\Beq
P_k(\tau)P_k^{\dagger}(\tau)={\rm I}_{4}\,.
\Eeq
We then impose that
\Beq
P_k(\tau)\begin{pmatrix}{\bf E}^u & {\bf F}^{\dagger} \\ {\bf F} & {\bf E}^v\end{pmatrix}P_k^{\dagger}(\tau)\,,
\Eeq
is diagonal, implying that the eigenvectors of $\begin{pmatrix}{\bf E}^u & {\bf F}^{\dagger} \\ {\bf F} & {\bf E}^v\end{pmatrix}$ are the columns of $P_k^{\dagger}(\tau)$.\\

Due to the nature of the equations of motion in the $S_-$ we can not have analytical expressions for the effective frequency ${\omega}^-_{s,k}(\tau)$ as opposed to Eq. \eqref{eq:omegagenpls} in $S_+$. Therefore, the expectation values of observables are calculated numerically.\\

At early times all off-diagonal terms of the Hamiltonian should vanish for both $s= \pm$. In other words $P_k(\tau) = {\bf I}_{4}$, which is equivalent to having
\Beq
\label{eq:HminsBD}
H_- = \begin{pmatrix} {\bf I}_{2} & 0\\  0& -{\bf I}_{ 2}  \end{pmatrix}\,.
\Eeq
It follows from equation (\ref{eq:EFm}) that 
\Beq
\lim\limits_{k\tau\to-\infty} F_s=0\,,
\lim\limits_{k\tau\to-\infty} E^{u}_{mix}=0\,,
\lim\limits_{k\tau\to-\infty} F_{s,mix}=0\,.
\Eeq

The above conditions are met, if the following equations are satisfied: \\ 
\begin{enumerate}

\item $\lim\limits_{k\tau\to-\infty}(u^{\uparrow}_{s,+}(k,\tau)-v^{\uparrow*}_{s,+}(k,\tau))=0$ ,$\lim\limits_{k\tau\to-\infty}(u^{\downarrow}_{s,+}(k,\tau)+v^{\downarrow*}_{s,+}(k,\tau))=0$. \\ It can be shown that these conditions are preserved by the equations of motion, so if imposed once they hold for any arbitrary $\tau$ as well
 \Beq
u^{\uparrow}_{s,+}(k,\tau)=v^{\uparrow*}_{s,+}(k,\tau)\,,  u^{\downarrow}_{s,+}(k,\tau)=-v^{\downarrow*}_{s,+}(k,\tau)\,. 
 \Eeq
\item $\lim\limits_{k\tau\to-\infty}\Re\big(u^{\uparrow*}_{s,+}(k,\tau)u^{\downarrow}_{s,+}(k,\tau)\big)=-1$ and $\lim\limits_{k\tau\to-\infty}\Re\big(v^{\uparrow*}_{s,+}(k,\tau)v^{\downarrow}_{s,+}(k,\tau)\big)=1$. This condition comes from Eq. \eqref{eq:HminsBD}. \\
\item $\lim\limits_{k\tau\to-\infty}u^{\uparrow, \downarrow}_{s,-}(k,\tau) = 0$ and $\lim\limits_{k\tau\to-\infty}v^{\uparrow, \downarrow}_{s,-}(k,\tau) = 0$. This condition is imposed so that the particle occupation number vanishes at early times.
Note that this is an arbitrary choice and one can make the mode functions vanish for, e.g., $p=+$.\\

\end{enumerate}
Using the above we can solve the equations of motion numerically.

\bibliographystyle{JHEP}
\bibliography{mybib}

\end{document}